\documentclass[final,epjc3]{svjour3}  
\smartqed  % flush right qed marks, e.g. at end of proof
\usepackage{amssymb}
\usepackage{amsmath}
\usepackage{graphicx}
\usepackage{dcolumn}
\usepackage{bm}
\usepackage{amssymb}
\RequirePackage{color}
\usepackage{datetime}
\usepackage{wasysym}
\usepackage[mathscr]{euscript}
\usepackage{hyperref}
\usepackage{slashed}
\usepackage{cancel}
\usepackage{ulem}
\usepackage[utf8]{inputenc} 
\newcommand{\bal}{\begin{align}}
\newcommand{\eal}{\end{align}}
\newcommand{\beq}{\begin{eqnarray}}
\newcommand{\eeq}{\end{eqnarray}}
\newcommand{\nnn}{\nonumber}
\newcommand{\nn}{\nonumber \\}
\newcommand{\es}{& = &}
\newcommand{\rs}{\, = \,}
\newcommand{\ps}{& + &}
\newcommand{\ms}{& - &}
\newcommand{\ts}{& \times &}
\newcommand{\nt}{\nn \ts}
\newcommand{\np}{\nn \ps}
\newcommand{\nm}{\nn \ms}

\newcommand{\cM}{ {\cal M} }
\newcommand{\cH}{ {\cal H} }

\newcommand{\cF}{ {\cal F} }
\newcommand{\cG}{ {\cal G} }

\newcommand{\cU}{ {\cal U} }
\newcommand{\cL}{ {\cal L} }

\newcommand{ \cS }{ {\text{$\mathcal{S}$}} } % ks

\newcommand{ \trZ }{ {\text{$\reflectbox{Z}$}} } % ks

\newcommand{\tdelta}{\tilde\delta}
\newcommand{\pd}{ {\partial} }
\newcommand{\ket}[1]{ {\left|{#1}\right\rangle} }
\newcommand{\bra}[1]{ {\left\langle{#1}\right|} }

\newcommand{ \up  }{ {\uparrow} }
\newcommand{ \down  }{ {\downarrow} }

\journalname{Eur. Phys. J. C}
\begin{document}

\title{Approximate Hamiltonian for baryons in heavy-flavor QCD%\thanksref{t1}
}
%\subtitle{Do you have a subtitle?\\ If so, write it here}

%\titlerunning{Short form of title}        % if too long for running head

\author{Kamil Serafin\thanksref{e1,fuw}
        \and
        Mar{\'i}a G{\'o}mez-Rocha\thanksref{e2,graz,granada}
        \and
        Jai More\thanksref{e3,iitb}
        \and
        Stanis{\l}aw D. G{\l}azek\thanksref{e4,fuw}
}

%\thankstext{t1}{Grants or other notes
%about the article that should go on the front page should be
%placed here. General acknowledgments should be placed at the end of the article.
\thankstext{e1}{e-mail: Kamil.Serafin@fuw.edu.pl}
\thankstext{e2}{e-mail: mgomezrocha@ugr.es}
\thankstext{e3}{e-mail: more.physics@gmail.com}
\thankstext{e4}{e-mail: Stanislaw.Glazek@fuw.edu.pl}

%\authorrunning{Short form of author list} % if too long for running head

\institute{Institute of Theoretical Physics,
           Faculty of Physics, 
           University of Warsaw,
           Pasteura 5, 02-093 Warsaw, Poland \label{fuw}
           \and
           Institute of Physics,
           University of Graz, NAWI Graz,
           A-8010 Graz, Austria \label{graz}
           \and
           Departamento de F\'isica At\'omica,
           Molecular y Nuclear and Instituto Carlos I de F\'isica
           Te\'orica y Computacional,
           Universidad  de  Granada, E-18071 Granada, Spain \label{granada}
           \and
           Department of Physics,
           Indian Institute of Technology Bombay,
           Powai, Mumbai 400076, India \label{iitb}
}

\date{Received: date / Accepted: date}
% The correct dates will be entered by the editor

\maketitle

\begin{abstract}

Aiming at relativistic description of gluons in hadrons,
the renormalization group procedure for effective particles
(RGPEP) is applied to baryons in QCD of heavy quarks. The
baryon eigenvalue problem is posed using the Fock-space
Hamiltonian operator obtained by solving the RGPEP equations
up to second order in powers of the coupling constant. The
eigenstate components that contain three quarks and two or
more gluons are heuristically removed at the price of inserting
a gluon-mass term in the component with one gluon. The resulting
problem is reduced to the equivalent one for the component of
three quarks and no gluons. Each of the three quark-quark interaction
terms thus obtained consists of a spin-dependent Coulomb term
and a spin-independent harmonic oscillator term. Quark masses
are chosen to fit the lightest spin-one quarkonia masses most
accurately. The resulting estimates for bbb and ccc states match
estimates obtained in lattice QCD and in quark models. Masses
of ccb and bbc states are also estimated. The corresponding
wave functions are invariant with respect to boosts. In the ccb
states, charm quarks tend to form diquarks. The accuracy of our
approximate Hamiltonian can be estimated through comparison 
by including components with two gluons within the same method. 
\keywords{QCD \and Heavy quarks \and Front Form \and Hamiltonian 
\and Renormalization \and Gluon mass \and Baryons}
% \PACS{PACS code1 \and PACS code2 \and more}
% \subclass{MSC code1 \and MSC code2 \and more}
\end{abstract}

%%%%%%%%%%%
 \section{ Introduction }
 \label{sec:intro}
%%%%%%%%%%%

Theoretically precise and phenomenologically accurate description 
of triply-heavy baryons as quantum states of quarks and gluons 
requires the formulation of QCD that satisfies several conditions. 
It ought to include a construction of the theory ground state, 
or vacuum, whose excitations are the quanta of quark and gluon 
fields. Since the canonical QCD Hamiltonian involves singularities 
and requires regularization, the theoretical formulation should 
include a mathematically precise renormalization procedure with 
clear physical interpretation, as a foundation of its predictability. 
The condition that individually quarks and gluons are not observed, 
implies that the formulation should allow for inclusion of 
confinement. The fact that heavy baryons may participate in 
processes whose description involves motion with speeds close 
to the speed of light, forces the formulation to be relativistic. 
In particular, it must guarantee description of baryons that have 
energies very much larger than their masses. Finally, knowing 
technical complexity of QCD and realizing that exact solutions 
are unlikely, it is necessary to demand that the formulation 
includes an outline of a process of successive approximations 
that stand a chance of systematically improving precision and 
accuracy of approximate solutions for observables. This article 
concerns a pilot application of an approach to heavy-quark QCD 
that in principle satisfies these requirements. 

Quark model represented baryons as bound states of three quarks, 
{\it e.g.} see~\cite{Capstick:2000qj,Capstick:1986bm}. In QCD, 
baryons are instead superpositions of states of quanta of quark 
and gluon fields. {\it A priori}, the number of quanta varies 
from three to infinity, across an infinite set of components. 
These quanta may have momenta ranging from zero to infinity. 
Their interactions diverge with their momenta. This article 
contributes to a development of a Hamiltonian approach to QCD 
that appears capable of filling the gap between the complex 
quantum-field picture and simple quark-model picture for 
hadrons~\cite{QQbarRGPEP}. Most succinctly, we illustrate a 
new method for solving the bound-state problem in canonical 
quantum field theories with asymptotic freedom in terms of 
its first application to the case of baryons in heavy-flavor QCD. 
Our method involves three consecutive steps: we solve our 
renormalization group equation for the front form (FF) Hamiltonian 
of the theory using the concept of effective particles in the 
Fock space; we reduce the resulting heavy-baryon eigenvalue 
problem for low-mass eigenstates to the eigenvalue problem 
solely for their Fock component of three effective quarks, 
using a gluon mass ansatz to account for the Fock components 
with more effective gluons than one; and we draw a qualitative 
sketch of the estimated low-mass heavy baryons spectrum that 
follows from the dominant mechanism of binding, while spin and 
other relatively small corrections to the effects of dominant 
interactions require future more elaborate calculations of 
higher-order using the same method. The theoretical challenge 
our method thus addresses is how to represent states of heavy 
baryons in terms of the Fock-space wave-functions for quarks 
and gluons that are invariant with respect to Lorentz boosts. 
The new results that our pilot study yields for heavy baryons,
including the approximate analytic formulae for their mass 
eigenvalues and corresponding boost-invariant wave functions 
that can be used in phenomenology of their production and 
detection, are thus derived in full detail from the heavy-flavor 
QCD supplied with our gluon mass ansatz. However, our pilot study of  
the low-mass triply heavy baryons involves severe simplifications. 
Similar simplifications are made in other approaches but without 
using the concept of effective particles that we introduce. 
In our approach, all the simplifications we make in the pilot 
study can be systematically removed within the same method while 
increasing its precision, as will be explained later on, but we 
do not address the question if the RGPEP may be used to derive the 
pNRQCD, which would require comparison of the dimensional regularization 
renormalization group equations with equations of the RGPEP, see 
the pertinent footnote on p. 456 in Ref.~\cite{tHooft:1973mfk}. 

We limit the theory to quarks that have masses much 
greater than $\Lambda_{QCD}$, excluding the top quark, 
and we consider the weak coupling limit~\cite{Wilsonetal}.
Creation of quark-antiquark pairs is neglected. Components 
with more than one gluon are eliminated, by assuming that their 
dominant effect in the component with one gluon is that the 
gluon has a mass, allowed to be a function of the gluon kinematic 
relative momentum with respect to the quarks. We use second-order 
perturbation theory to derive the resulting effective Hamiltonian 
for baryons that only acts in the component with three quarks. 
We compare the quark-quark interaction terms in this Hamiltonian to 
similar terms in the Hamiltonian that only acts in the quark-anti-quark 
component in quarkonia, previously derived using the same 
method~\cite{QQbarRGPEP}. Masses of heavy baryons are 
estimated by solving the resulting eigenvalue equation in the 
nonrelativistic limit. The parameters involved (the running 
coupling and the quark masses) are chosen using heavy quarkonia 
experimental data. In this way, our estimates for baryon masses 
contain no new parameters. More precisely, the coupling constant 
is extrapolated from a formally infinitesimal value of weak-coupling 
limit to the value implied at the quark-mass scale by the known 
coupling constant at the scale of $Z$-boson mass. Quark masses 
are adjusted to the known spectra of heavy quarkonia. The scale 
parameter for hadrons built from different flavors is fixed by 
a linear interpolation between its one-flavor values. This will 
be explained in detail later. 

The concept of effective-gluon mass that we use is explained 
in Sec.~\ref{sec:mass}, followed by a brief outline of our 
method in Sec.~\ref{sec:RGPEP}. The method is called the
renormalization group procedure for effective particles (RGPEP). 
Our concept of the gluon mass differs from the concepts discussed 
in the literature, see, {\it e.g.,} 
Refs.~\cite{Parisi:1980jy,Cornwall:1981zr,Aguilar:2017dco}. 
The second-order baryon eigenvalue problem is outlined in 
Sec.~\ref{sec:baryonEVE}. Details of the effective quark-quark 
interaction terms in $ccc$ and $bbb$ baryons, implied by the gluon 
mass, are described in Sec.~\ref{sec:baryonOSC}, including 
a comparison with the case of heavy quarkonia. Sec.~\ref{sec:twoflavors} 
extends the calculation to the $ccb$ and $bbc$ baryons. The 
resulting estimates for baryon masses are described in 
Sec.~\ref{sec:sketch}. Comments concerning the RGPEP 
calculation of effective Hamiltonians in QCD in orders higher 
than second and beyond perturbation theory conclude the 
paper in Sec.~\ref{sec:conclusion}. Details of our fit to the 
spectra of quarkonia are described in \ref{app:fits}. 
Values of the RGPEP scale parameter we use are listed
in \ref{app:predictions}. \ref{app:frequencies} discusses 
dependence of harmonic oscillator frequencies on the scale 
parameter. \ref{app:wavefunctions} provides a detailed description 
of the baryon wave functions that are used in our estimates 
and \ref{app:masses} presents explicit formulas for the 
associated heavy-baryon masses.

%%%%%%%%%%%%%%%%%%
 \section{ Assumption of gluon mass }
 \label{sec:mass}
%%%%%%%%%%%%%%%%%%

Theoretically, a baryon state in heavy-flavor QCD is a superposition 
of states of virtual, point-like quarks and gluons, 
\beq
\ket{\Psi} \es \ket{3Q} + \ket{3Q \, G} + \ket{3Q \, 2G} + \dots \  .
\eeq
Components that include quark-anti-quark pairs are considered 
very small because quarks are heavy. In contrast, components 
with gluons are included because in canonical QCD gluons are 
massless. However, using the massless gluons in the expansion 
and limiting their number in a computation one expects to obtain 
the spectrum of excited baryons that gets dense toward the
free quark threshold. The same feature is expected to occur in 
such computations of spectrum of quarkonia. Physically, the 
latter is observed to be not dense~\cite{PDG2016,Eichten:1978tg}. 
For example, the $s$-wave $c \bar c$ or $b \bar b$ mass splittings 
are on the order of half GeV and they do not decrease with the 
excitation number as they would if the interaction was purely of 
the Coulomb type, like in QED with massless photons. The mass 
splittings in the heavy baryon spectrum are also not expected to 
rapidly decrease with the excitation number. To describe physical 
splittings, excitations of the gluon field must involve considerable 
energy. This requirement can be addressed using the concept of 
a gluon mass~\cite{QQbarRGPEP}. 

Introduction of a mass term for gluons in the canonical 
Hamiltonian of QCD would spoil its gauge-theory structure. 
Instead, we introduce a gluon mass in solving the eigenvalue 
problem of a Hamiltonian $H_t$ that is derived using the 
RGPEP~\cite{pRGPEP}, see below. The parameter $t$ is the 
renormalization group parameter. It is useful to think about
it as $t=s^4$, where $s$ has an interpretation of the size of 
effective particles. Note that the effective-particle size $s$, 
as a parameter of renormalization group procedure in 
quantum field theory, is not mathematically related to the 
phenomenological size-parameters for gluons, such as, for 
example, in Ref.~\cite{Semay:1997ys}. Canonical gluons are 
considered point-like. Instead of using canonical gluons, the 
goal is to represent a heavy baryon by a superposition 
\beq
\ket{\Psi} \es \ket{3Q_t} + \ket{3Q_t \, G_t}
+ \ket{3Q_t \, 2G_t} + \dots \  ,
\label{psit}
\eeq
where the quarks and gluons are the effective particles of 
size $s$. We introduce the gluon mass in the effective QCD 
eigenvalue problem for heavy baryons within the same 
computational scheme that we previously applied to heavy 
quarkonia~\cite{QQbarRGPEP}.\footnote{Note that the 
gluon mass is introduced not in the Lagrangian or canonical 
QCD Hamiltonian, but as a candidate for an approximation in 
solving the eigenvalue equation. Thus, the ansatz does not 
violate the gauge symmetry in the canonical theory.} 

Our leading principle is that the gluon mass $\mu_t$ is the 
minimal price we have to pay for limiting the expansion in 
Eq.~(\ref{psit}) to the first two terms. Such limitation makes 
sense because interactions in the Hamiltonian $H_t$ contain 
vertex form factors. The form factors are obtained by solving 
the RGPEP Eq.~(\ref{RGPEP}), with the initial condition 
provided by the canonical QCD Hamiltonian with regularization 
and counterterms. These form factors cause that interactions 
cannot change invariant masses of component states by 
amounts exceeding $1/s$.  Therefore, the effective gluons 
of size $s$ cannot be as copiously produced as the point-like 
gluons can in the canonical representation of QCD. It is  
plausible that inclusion of a few effective components is 
sufficient to accurately describe a heavy-baryon solution. 
Regarding attempts of relating our effective gluon quanta 
to gluon field degrees of freedom in other approaches, 
it would be of general interest to find out if lattice 
studies, such as in Refs.~\cite{Juge:2002br,Takahashi:2002it}, 
can introduce interpolating operators that are capable of 
identifying properties of the same degrees of freedom.

Although the number of gluons $G_t$ that need to be included 
in the effective representation of a low-mass solution to the 
QCD eigenvalue problem is expected to be limited, direct 
inspection shows that inclusion of even a few components 
still leads to a mathematically difficult equations for their
coupled-channel dynamics. The results presented in this paper 
follow from the effective eigenvalue problem that is obtained 
by using the hypothesis that the contribution of all components 
other than $|3Q_t\rangle$ and $|3 Q_t \, G_t \rangle$ may be 
approximated by inclusion of a gluon mass, $\mu_t$ for 
the gluon in component $|3Q_t \,  G_t \rangle$. 

The mass assumption is falsifiable by extending the calculation
to explicitly include more components and relegating the gluon 
mass ansatz to states with more gluons than one. The purpose
would be to verify if the finite value of the RGPEP parameter $s$ 
on the order of quark Compton wave-length is sufficient to 
prevent the spill of probability to states with many gluons,
especially when the coupling constant is small. The latter 
situation is expected to occur for the quark masses that are 
much greater than $\Lambda_{QCD}$. This is precisely the 
reason for us to study the dynamics of gluons using the RGPEP 
first in the context of heavy-flavor QCD. In order to simplify 
the problem and thus increase a chance of understanding the 
dynamics of effective gluons whose masses are likely to be 
much larger than $\Lambda_{QCD}$, we exclude from the 
theory the quarks that have masses much smaller than 
$\Lambda_{QCD}$. If the latter were included in the theory, 
they could appear in large numbers in the effective Fock-space 
basis and complicate the dynamics, as massless gluons do.

%%%%%%%%%%%%%%
 \section{ RGPEP for hadrons }
 \label{sec:RGPEP}
%%%%%%%%%%%%%%

The RGPEP provides equations for calculating the re\-normalized 
Hamiltonian $H_t$ from the canonical one that includes 
regularization and counterterms. It is also used to calculate the 
counterterms. Eigenstates of $H_t$ define hadrons in terms of 
effective particle basis in the Fock space. We first consider QCD 
of only one flavor of heavy quarks, useful in discussing dynamics
in baryons made of quarks of one flavor. The case of baryons 
made of two types of heavy quarks is discussed in Sec.~\ref{sec:twoflavors}. 
We calculate $H_t$ using expansion in powers of a formally
infinitesimal coupling constant, up to terms of second-order. 
Results of our second-order calculations are later compared
with results obtained in quark models and in lattice approach
to QCD. 
      
%%%%%%%%%%%%%%%%%%%%%%%%%%%%%%%%%%%%%%%%%%%%%%%%
 \subsection{ Canonical Hamiltonian }
%%%%%%%%%%%%%%%%%%%%%%%%%%%%%%%%%%%%%%%%%%%%%%%%

The Lagrangian for one-flavor QCD is
\beq
\cL
\es
  \bar \psi (i\slashed D - m)\psi
- \frac{1}{2} \text{tr} F^{\mu\nu}F_{\mu\nu}
\ .
\eeq
We use the FF of Hamiltonian 
dynamics~\cite{Dirac1949} and employ canonical 
quantization to derive the corresponding Hamiltonian 
$\hat H_{QCD}^{\rm can}$ in the gauge $A^+=0$,
\beq
\hat H_{QCD}^{\rm can} \rs \hat P^-
\rs
\int dx^- d^2 x^\perp \  : \hat {\cal H}_{x^+=0} :
\ .
\eeq
We adopt the FF notation of Refs.~\cite{Lepage:1980fj,Brodsky-Pauli-Pinsky}. 
The Hamiltonian operator density, $: \hat {\cal H}_{x^+=0} :$ 
integrated over the front $x^+=0$, is expressed in terms of the 
quantum fields
\begin{align}
\hat\psi
\rs
\sum_{\sigma c } \int[p]
\left[  \chi_c u_{p\sigma} \hat b_{p\sigma c } e^{-ipx}
      + \chi_c v_{p\sigma} \hat d^\dagger_{p\sigma c } e^{ipx}
\right]_{x^+=0}
\ ,
\\
\hat A^\mu
\rs
\sum_{\sigma c} \int[p]
\left[  { T^c} \varepsilon^\mu_{p\sigma} \hat a_{p\sigma c} e^{-ipx}
      + T^c \varepsilon^{\mu *}_{p\sigma} \hat a^\dagger_{p\sigma c} e^{ipx}
\right]_{x^+=0}
\ ,
\end{align}
where $\int[p]=\int_0^\infty dp^+ \int d^2 p^\perp /[2p^+(2\pi)^3]$,
$u_{p\sigma}$ and $v_{p\sigma}$ are the Dirac spinors,
$\varepsilon^\mu_{p\sigma}$ is the transverse-gluon polarization 
vector, $\chi_c$ and $T^c$ denote three-component color vector 
for quarks and eight-component color matrix vector for gluons, 
while $\sigma$ and $c$ stand for their spins and colors, respectively. 
We omit the hats and normal ordering symbols in further formulas. 
In second-order calculation, only two interaction terms count, the 
quark-gluon interaction,
\beq
\cH_{\psi A\psi}
\es
A_\mu^a
\ j_{\rm quark}^{a \mu}  
\ ,
\eeq
and the instantaneous quark-quark interaction,
\beq
\cH_{(\psi\psi)^2}
\es
\frac{1}{2}
j_{\rm quark}^{a +}  
\ \frac{1}{(i\pd^+)^2} \ 
j_{\rm quark}^{a +}  
\ ,
\label{cHpsipsi2}
\eeq
where
\beq
j_{\rm quark}^{a \mu}  
\es  
g_{\rm bare} \ \bar \psi \gamma^\mu T^a \psi
\ .
\eeq

%%%%%%%%%%%%%
\subsection{Regularization}
%%%%%%%%%%%%%

$H_{QCD}^{\rm can}$ is regularized by inserting 
cutoff functions $r_{21.3}$ and $r_{C\,12.1'2'}$ 
in the interaction vertices, as shown below and 
further explained in Appendix~A of Ref.~\cite{QQbarRGPEP}. 
The terms that contribute to the baryon problem are 
\begin{eqnarray}
H^{\rm can \ R}_{\rm QCD} 
\es 
H_{\rm free} + g_{\rm bare} H_{1}^R + g_{\rm bare}^2 H_{QQ\, \text{inst}}^R  \  
.
\end{eqnarray}
The \textit{free}, or kinetic term is
\begin{align}
H_{\rm free}
\rs
\sum_{\sigma c} \int[p] E_{q} \,
b_{p\sigma c}^\dag b_{p\sigma c}
+
\sum_{\sigma c} \int[p] E_{g} \,
a_{p\sigma c}^\dag a_{p\sigma c}
\ ,
\end{align}
where  $E_{q}$ and $E_{g}$ are the FF quark and 
gluon energies, respectively. In the quark-gluon 
vertex,
\beq
H_{1}^R
\es
\int_{123} r_{21.3}
B_{21.3}
\ b_{2}^\dag a_{1}^\dag b_{3}
+ \text{h.c.} \ ,
\label{Bare1st}
\eeq
the first term corresponds to emission and the second 
to absorption of a gluon by a quark. Numbers 1, 2, 3 
stand for sets of quantum numbers of particles 1, 2 
and 3, and $\int_{123}$ includes integration over
momenta and summation over spins and colors of particles
$1$, $2$ and $3$. In the factor
\beq
B_{21.3}
\es
  \tdelta_{21.3}
\ t^1_{23}
\ \bar u_2 \slashed \varepsilon_1^* u_3 \ ,
\eeq
$t^1_{23} = \chi_{c_2}^\dag T^{c_1} \chi_{c_3}$. 
The tilde over $\delta$ indicates an implicit factor $2(2\pi)^3$,
multiplying the Dirac $\delta$-function of momentum conservation. 
The function $r_{21.3}$ cuts off the large relative transverse 
momenta and small fractions of plus momenta for the particles 
involved in the vertex, {\it cf.} Appendix~A in~\cite{QQbarRGPEP}.  
The instantaneous interaction of Eq.~(\ref{cHpsipsi2}) yields
\beq
H_{QQ\, \text{inst}}^R
\es
  2 
  \int_{121'2'}
  \tilde \delta_{12.1'2'}
  \sqrt{x_{1} x_{2} x_{1'} x_{2'}}
  \ \delta_{\sigma_1\sigma_{1'}} \delta_{\sigma_2\sigma_{2'}}
% \nt
  \frac{r_{C\,12.1'2'}}
  {(x_{1}-x_{1'})^2}\, 
  t^a_{11'} t^a_{22'}
  \ b_{1}^\dagger b_{2}^\dagger b_{2'} b_{1'}
\ .
\nn
\eeq

It should be noted that the cutoff functions we use in the 
interaction terms imply that quanta with small plus momentum 
cannot participate in the dynamics obtained in a perturbative 
solution to the RGPEP equation of a finite order. This is 
important because the resulting dynamics does not involve 
quantum field modes that in the instant form are associated 
with the vacuum state~\cite{Wilsonetal}. The mechanism is 
the same as in the similarity renormalization group 
procedure~\cite{SRG1}. Therefore, also similarly, the physical 
effects associated with the vacuum in the instant form of dynamics 
are expected to appear only as new Hamiltonian interaction terms 
in the FF of dynamics. Thus, the RGPEP approach allows 
one to circumvent the need for finding the vacuum state and 
instead offers the possibility of finding interactions 
corresponding to the unknown state. Since the interactions 
act on the field quanta that are hadronic constituents, the 
vacuum effects can be thought of as limited to the hadronic 
interior, though they have a universal origin for all 
hadrons~\cite{Casher:1974xd,Maris:1997hd,Brodsky:2010xf,GlazekCondensatesAPP,Brodsky:2012ku}. 

%%%%%%%%%%%%%%%%%%%%%%%%%%%%%%%%%%%%%%%%%%%%%%%%
 \subsection{ Renormalized Hamiltonian }
%%%%%%%%%%%%%%%%%%%%%%%%%%%%%%%%%%%%%%%%%%%%%%%%

We call the regularized canonical Hamiltonian 
for quanta of size $s=0$ the \textit{initial Hamiltonian},
since it provides an initial condition for solving the
RGPEP equation. Its solution defines a family of 
renormalized Hamiltonians $H_t$, which are written in
terms of the operators $q_t$ that create or annihilate 
particles of size $s>0$, $t=s^4$. The latter operators 
are defined by means of a unitary transformation $\cU_t$,
\beq
q_t \es \cU_t \, q_0 \, \cU_t^\dagger
\ .
\label{qs}
\eeq
The idea is that nonzero size eliminates divergent
integrals. Hence, the effective Hamiltonians cannot 
be sensitive to the cutoff parameters in the cutoff
functions. This implies that the regularized canonical 
Hamiltonian needs to be supplemented with counter-terms, 
which ensures that the renormalized Hamiltonians do
not depend on the regularization. The problem is to
define $\cU_t$ that generates Hamiltonians $H_t \equiv 
H_t(q_t)$ in a suitable operator basis. Instead of directly 
defining $\cU_t$, we define $\cH_t = H_t(q_0)$, in 
which the products of operators from the bare theory 
have the coefficients from renormalized theory. By 
definition, it obeys 
\beq 
\cH'_t \es
[ \cG_t , \cH_t ] \ ,
\label{RGPEP}
\eeq 
where prime denotes derivative with respect to the 
parameter $t$. $\cG_t$ is called a generator of the 
RGPEP. It is set to 
\beq
\cG_t \es [ \cH_{\rm free}, \tilde \cH_t ] \ ,
\label{RGPEPgenerator}
\eeq
where $\cH_{\rm free}$ is the free part of $\cH_t$, and is
identical with $H_{\rm free}$. The tilde above $\cH_t$ means
that coefficients in front of interaction terms are multiplied 
by the square of total $+$-momentum entering the vertex. 
The RGPEP design guarantees that the interaction vertices 
that change invariant mass of interacting particles by more 
than $1/s$ are exponentially suppressed~\cite{pRGPEP}, 
{\it cf.} Eq.~(\ref{Ht1}).

In the present work, we solve Eq.~(\ref{RGPEP}) using 
expansion in powers of renormalized coupling constant 
$g_t$, up to second order,
\beq
H_t
\es
  H_{t0}
+ g_t H_{t1}
+ g_t^2 H_{t2\, QQ}
+ g_t^2 H_{t2\, \delta m}
\ .
\eeq
The zero-order term, $H_{t0}$, corresponds to terms 
without running coupling. The only difference between 
$H_{t0}$ and the bare expression $H_\text{free}$ is 
the presence of effective creation and annihilation 
operators in place of bare ones,
\begin{align}
H_{t0}
\rs
\sum_{\sigma c} \int[p] E_{q} \,
b_{t\,p \sigma c}^\dag b_{t\,p \sigma c}
+
\sum_{\sigma c} \int[p] E_{g} \,
a_{t\,p \sigma c}^\dag a_{t\,p \sigma c}
\ .
\end{align}
The solution for quark-gluon interaction term of 
order $g_t$, apart from substitution of bare operators 
by effective ones, differs from the bare theory term 
of Eq.~(\ref{Bare1st}) by the presence of form factors 
$f_{t\,21.3}$. Namely,
\beq
H_{t1}
\es
\int_{123} r_{21.3} f_{t\,21.3}
B_{21.3}
\ b_{t\,2}^\dag a_{t\,1}^\dag b_{t\,3}
+ \text{h.c.}
\ ,
\label{Ht1}
\eeq
with 
\begin{eqnarray}
f_{t\,21.3} \es 
\exp[-(\cM^2_{21} - m^2)^2\, t] \ ,
\end{eqnarray}
where $\cM^2_{21}$ is the square of free invariant mass
of particles $1$ and $2$. The second-order terms that matter 
are the quark-quark interaction and quark self-interaction terms.
The quark-quark interaction, $H_{t2\, QQ}$ is a sum of two
parts. The one that stems from the instantaneous interaction,
\beq
H_{t2\,QQ\,\text{inst}}
\es
  2 
  \int_{121'2'}
  \ \tilde \delta_{12.1'2'} \ 
  \sqrt{x_{1} x_{2} x_{1'} x_{2'}}
  \ \delta_{\sigma_1\sigma_{1'}} \delta_{\sigma_2\sigma_{2'}}
\nt
  f_{t\,12.1'2'}
  \frac{r_{C\,12.1'2'}}
  {(x_{1}-x_{1'})^2}
  \, t^a_{11'} t^a_{22'}
  \, b_{t\,1}^\dagger b_{t\,2}^\dagger b_{t\,2'} b_{t\,1'}
\ ,
\eeq
differs from $H_{QQ\, \text{inst}}$ by effective 
particle operators and a form factor~\cite{QQbarRGPEP}
\begin{eqnarray}
f_{t\,12.1'2'} 
\es
\exp[-(\cM^2_{12} - \cM^2_{1'2'})^2 \, t] \ .
\end{eqnarray}
The other one that stems from exchange of transverse gluons,
\beq
H_{t2\,QQ\,\text{exch}} 
\es
 \frac{1}{2}
  \int_{121'2'} f_{12.1'2'}\ \tdelta_{12.1'2'}
  \ t^a_{11'} t^a_{22'}
  \frac{ d_{\mu\nu}(p_4) }{ p_4^+ }
  \ j^\mu_{11'} j^\nu_{22'}
  \ b_{t\,1}^\dag b_{t\,2}^\dag b_{t\,2'} b_{t\,1'}
\nt
\left[
  \ \theta(z_{11'}) F_{\trZ}(12;1'2')
  \, r_{1'4.1} r_{24.2'}
% \right.
% \left.
+
  \theta(-z_{11'}) F_{Z}(12;1'2')
  \, r_{2'4.2} r_{14.1'}
\right]
\ , \nn
\label{exchRGPEP}
\eeq
where $j^\mu_{ij}=\bar u_{i} \gamma^\mu u_{j}$,
involves factors
\begin{align}
F_{\trZ}(12;1'2')
\rs
- \frac{ \left( p_1^+ \cS_{1'4} + p_{2'}^+ \cS_{24} \right)
         \left( 1 - \frac{ f_{t\,1'4.1} f_{t\,24.2'} }{ f_{t\,12.1'2'} } \right) }
       { \cS_{1'4}^2 + \cS_{24}^2 - (\cS_{12}-\cS_{1'2'})^2 }
\ ,
\\
F_{Z}(12;1'2')
\rs
- \frac{ \left( p_2^+ \cS_{2'4} + p_{1'}^+ \cS_{14} \right)
         \left( 1 - \frac{ f_{t\,14.1'} f_{t\,2'4.2} }{ f_{t\,12.1'2'} } \right) }
       { \cS_{2'4}^2 + \cS_{14}^2 - (\cS_{12}-\cS_{1'2'})^2 }
\ ,
\end{align}
with $\cS_{ij} = \cM_{ij}^2 - m_i^2$ and $m_i$ denoting
the mass of particle $i$. Moreover, the gluon momentum is
\beq
p_4^+ \es |p_1^+ - p_{1'}^+|
\ ,
\\
p_4^\perp
\es
\epsilon(z_{11'}) ( p_1^\perp - p_{1'}^\perp )
\ ,
\\
z_{11'}
\es
\frac{p_{1}^+ - p_{1'}^+}{p_{1}^+ + p_{2}^+}
\ ,
\eeq
where $\epsilon(z)$ denotes the sign of $z$.

The differences between Eq.~(\ref{exchRGPEP})
and the corresponding equations for quarkonia~\cite{QQbarRGPEP},
are: quark current $j_{22'}$ instead of anti-quark current
$\bar j_{2'2} = \bar v_{2'} \gamma^\mu v_{2}$, transposition
of matrix $t^4_{22'}$, overall
sign difference, color factor $-2/3$ instead of $4/3$ (when
acting on a color singlet state) and symmetrization factor
$1/2$ that is absent in quarkonia.

Renormalized quark self-interaction mass correction is 
\beq
H_{t2\,\delta m}
\rs
\int_{1}
\frac{ m_{t\,2}^2 }{p_1^+}
\ b^\dag_{t\,1} b_{t\,1}
\ ,
\eeq
where
\begin{align}
m^2_{t\,2}
\rs
  \frac{4}{3}
  \sum_{\sigma_5}
  \int[45]
  p_{1}^+\,
  \frac{e^{-2 t \cS_{54}^2}}{\cS_{54}}
  r_{54.1}^2
  \, \tilde \delta_{54.1}
  \, d_{\mu\nu}(p_4)
  \, j_{15}^\mu \, j_{51}^\nu
\ .
\end{align}

%%%%%%%%%%%%%%%%%%%%%%%%%%%%%%%%%%%%%%%%%%%%%%%%
 \subsection{ Bound-state eigenvalue problem }
%%%%%%%%%%%%%%%%%%%%%%%%%%%%%%%%%%%%%%%%%%%%%%%%

Thanks to asymptotic freedom~\cite{AF}, $g\equiv g_t$
is small in Hamiltonians with small $t$. This holds 
for $\lambda = 1/s$ much larger than the scale of 
$\Lambda_{\rm QCD}$ in the RGPEP scheme. We formally 
consider
\beq
m \gg \lambda = s^{-1} \gg \Lambda_{\rm QCD}
\label{scales}
\ ,
\eeq
which allows us to simplify the eigenvalue problem 
\beq
H_t \ket{\Psi}
\es 
 E \ket{\Psi} \ .
\label{eigenproblemHt}
\eeq
%\begin{widetext}
In the baryon eigenstates represented using 
Eq.~(\ref{psit}), we can neglect Fock sectors 
with more than three quarks, because of the 
first inequality in Eq.~(\ref{scales}). In a matrix 
form, the eigenvalue problem reads
\begin{equation}
\left\{
\left[ \begin{array}{lll}
       . & . & . \\  
       . & H_{t\,0} + g^2 H_{t2} &  g H_{t1}  \\
       . & g H_{t1} & H_{t\,0} + g^2 H_{t2}
       \end{array} \right]
-E 
\right\}
\left[  \begin{array}{l}
        .                    \\ 
        | 3Q_t \, G_t \rangle   \\
        | 3Q_t  \rangle 
        \end{array} \right]
= 0 \ ,  
\label{eigenproblemmatrix}
\end{equation}
%\end{widetext}
where $H_{t2} = H_{t2\, QQ} + H_{t2\, \delta m}$ and dots stand
for the Fock components with more than one effective gluon and
for the Hamiltonian terms that involve those components. 

%%%%%%%%%%%%%%%%%%%%%%%%%%%%%%%%%%%%%%%%%%%%%%%%
 \subsection{ Gluon mass ansatz in the effective eigenvalue problem }
%%%%%%%%%%%%%%%%%%%%%%%%%%%%%%%%%%%%%%%%%%%%%%%%

Similarly to the case of heavy quarkonia~\cite{QQbarRGPEP}, 
we remove the Fock components and Hamiltonian matrix elements 
that involve more than one gluon. The price for this removal is a 
gluon-mass ansatz for the component $| 3Q_t \, G_t\rangle$. 
Indeed, a gluon mass term appears when one uses Gaussian 
elimination to express the component $| 3Q_t \, 2 G_t\rangle$ in 
terms of component $| 3Q_t \, G_t\rangle$. Our working hypothesis 
is that other terms can also be dropped once a mass ansatz is 
introduced. The reduced eigenvalue problem with the ansatz 
mass term $\mu_t^2$ is
\begin{align}
\left\{
\left[ \begin{array}{ll}
       H_{t\,0} + \mu_t^2  &  g H_{t1}  \\  
       g H_{t1} & H_{t\,0} + g^2 H_{t2}
       \end{array} \right]
- E 
\right\}
\left[  \begin{array}{l} 
        | 3Q_t \, G_t \rangle   \\
        | 3Q_t  \rangle 
        \end{array} \right] 
\rs 0 \  .
\end{align}
This two-component problem is reduced to an equation 
for the component $|3Q_t\rangle$, using a transformation
$R$ succinctly described in~\cite{Wilson1970}. Up to 
terms order $g^2$ and using notation $r$ and $l$ for
right and left states in the matrix elements, one obtains
\beq
\bra{l} H_{\rm{eff}\,t} \ket{r}
\es
\bra{l} \left[
  H_{t0} + g^2 H_{t2}
\right.
\np
\left. 
  \frac{1}{2} g H_{t1}
\left(
    \frac{1}{E_l - H_{t0} - \mu_t^2}
  + \frac{1}{E_r - H_{t0} - \mu_t^2}
  \right) g H_{t1}
\right] \ket{r}
\ .
\label{Heff2nd}
\eeq

%%%%%%%%%%%%%%%%
 \section{ 3Q eigenvalue problem }
 \label{sec:baryonEVE}
%%%%%%%%%%%%%%%%

The three-quark component of a baryon satisfies
the FF eigenvalue equation
\beq
H_{{\rm eff}\,t}|3Q_t\rangle
\es
\frac{M^2 + (P^\perp)^2}{P^+} |3Q_t\rangle
\ ,
\eeq
in which the state $|3Q_t\rangle$ is defined by
\begin{align}
\ket{3Q_t}
\rs
\int_{123}
P^+\tdelta_{P.123}
\,\psi_t(123)
\,\frac{\epsilon^{c_1c_2c_3}}{\sqrt{6}}
b_{t\,1}^\dag b_{t\,2}^\dag b_{t\,3}^\dag
\ket{0}
\ .
\end{align}
The spin-momentum wave function, $\psi_t(123)$, is 
multiplied by the color factor $\epsilon^{c_1c_2c_3}/\sqrt{6}$.
The eigenvalue equation for the spin-momentum wave 
function reads
\begin{align}
&\left(
  \frac{\mathscr M^2_{1,\,t} + (p_1^\perp)^2}{p_1^+}
+ \frac{\mathscr M^2_{2,\,t} + (p_2^\perp)^2}{p_2^+}
+ \frac{\mathscr M^2_{3,\,t} + (p_3^\perp)^2}{p_3^+}
\right)
\psi_t(123)
\nn
& +
  g^2
  \sum_{\sigma_{1'}\sigma_{2'}} \int[1'2']
  \ \tdelta_{12.1'2'}
  \ U_{t\,\text{eff}}(12;1'2')
  \ \psi_t(1'2'3)
\nn
&
+
  g^2
  \sum_{\sigma_{3'}\sigma_{1'}} \int[3'1']
  \ \tdelta_{31.3'1'}
  \ U_{t\,\text{eff}}(31;3'1')
  \ \psi_t(1'23')
\nn
&
+
  g^2
  \sum_{\sigma_{2'}\sigma_{3'}} \int[2'3']
  \ \tdelta_{23.2'3'}
  \ U_{t\,\text{eff}}(23;2'3')
  \ \psi_t(12'3')
% \nn
% &
\rs
\frac{ M^2 + (P^\perp)^2 }{ P^+ }
\ \psi_t(123)
\ .
\label{QQQeffEVE}
\end{align}
\begin{figure}
\centering
\includegraphics[width=0.2\textwidth]{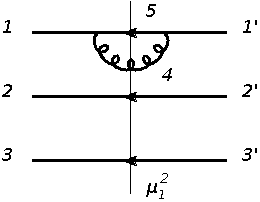}
\quad\quad
\includegraphics[width=0.2\textwidth]{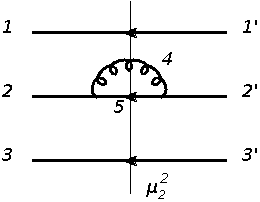}
\quad\quad
\includegraphics[width=0.2\textwidth]{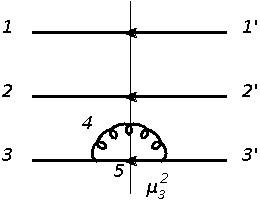}
\caption{Self-interaction of effective quarks.}
\label{fig:self}
\end{figure}
The first line contains the kinetic energy, including 
self-interaction terms illustrated in Fig.~\ref{fig:self},
with
\beq
\mathscr M^2_{i,\,t}
\es
  m^2
+ \frac{4}{3}
  g^2
  \int[x_{4/i}\kappa_{4/i}]
  \ r_{54.i}^2
  \frac{
  e^{-2 t \cS_{54}^2}
  }{
  \cS_{54}
  \cS_{54\,\mu_i}
  }
  \frac{ \mu^2_{i}}{x_{4/i}}
\nt
  \left[
  2\left( \frac{2}{x_{4/i}} - 2 + x_{4/i} \right)
  \cS_{54}
- 4 m^2
  \right]
\\
\es
m^2 + \frac{4}{3} 
g^2 \int \frac{ d^2 \kappa^{\perp } dx }{ 2(2\pi)^3 x ( 1 - x ) }
\, r_{54.i}^2  \,
e^{-2\left( \cM^2-m^2 \right)^2t}
\nt
\left[ 
\sum_{\sigma_5} (\bar u_{i} \gamma^\mu  u_5) 
\, (\bar u_5 \gamma_\mu u_{i})
+ ( m^2 - \cM^2 ) 
\, 4 \frac{ 1 - x }{ x }
\right] 
\nt
\left(
\frac{ 1 }{ m^2 - \cM^2 } 
 - \frac{ 1 }{ m^2 - \cM^2 - {\mu^2_{i} / x} }
\right)
\ ,
% & 
%%%%%%%%%%%%%%%%%%%%%%%%%%%
\eeq
where $x=x_{4/i}$, $\cM^2=\cM_{54}^2$, $\cS_{54\,\mu_i} =
\cS_{54} + \mu_i^2  x^{-1}_{4/i}$. The mass ansatz
$\mu^2$, which is allowed to be a function of the gluon
relative momentum with respect to the three quarks, yields
$\mu^2_i = \mu^2(p_4,p_5,p_j,p_k)$, so that $\mu^2_1 =
\mu^2(p_4,p_5,p_2,p_3)$, $\mu^2_2 = \mu^2(p_4,p_5,p_3,p_1)$
and $\mu^2_3 = \mu^2(p_4,p_5,p_1,p_2)$, see Fig.~\ref{fig:self}.

\begin{figure}
\centering
\includegraphics[width=0.2\textwidth]{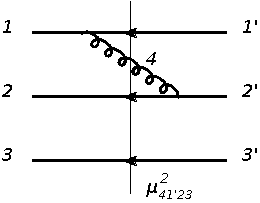}
\hspace{1em}
\includegraphics[width=0.2\textwidth]{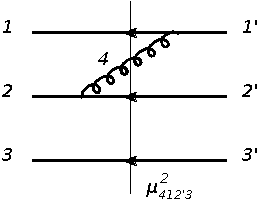}
\\
\includegraphics[width=0.2\textwidth]{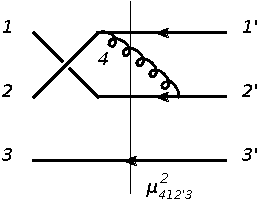}
\hspace{1em}
\includegraphics[width=0.2\textwidth]{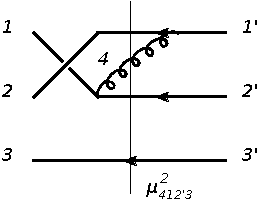}
\caption{Interaction of effective quarks by exchange 
of an effective gluon.}
\label{fig:exchange}
\end{figure}

The interactions include instantaneous and exchange terms
$U_{t\,\text{eff}} = H_\text{inst} + H_\text{exch}$,
\beq
H_\text{inst}(12;1'2')
\es
\text{S}
\left[
- \frac{2}{3}
  \ r_{C\,12.1'2'}
  \ f_{t\,12.1'2'}
  \frac{j^+_{11'} j^+_{22'}}{(p^+_{4})^2}
\right]
\ ,
\eeq
\beq
&&
H_\text{exch}(12;1'2')
\rs
\text{S}
\left\{
- \frac{2}{3}
  \ \frac{ d_{\mu\nu}(p_4) }{ p_4^+ }
  \ j^\mu_{11'} j^\nu_{22'}
\right.
\nn &&
\left.
\phantom{\frac{{}_{}}{{}^{}_{}}}
\times \hspace{-0.1em}
\left[
  \theta(z_{11'}) r_{1'4.1} r_{24.2'}
  \, \cF_\trZ^{12}
+ \theta(-z_{11'}) r_{2'4.2} r_{14.1'}
  \, \cF_Z^{12}
\right]
\right\}
\ ,
\nn
\eeq
where $\text{S}$ denotes symmetrization $1 \leftrightarrow 2$,
or, symbolically, $(12 + 21)/2$, see Fig.~\ref{fig:exchange}. 
One obtains
\begin{align}
\cF_\trZ^{12}
&=
  f_{12.1'2'} \ F_{\trZ}(12;1'2')
+ f_{1'4.1} f_{24.2'} \ R_\trZ(12;1'2')
\ ,
\\
\cF_Z^{12}
&=
  f_{12.1'2'} \ F_{Z}(12;1'2')
+ f_{2'4.2} f_{14.1'} \ R_Z(12;1'2')
\ ,
\end{align}
\begin{align}
R_\trZ(12;1'2')
&=
\frac{ -p_{1}^+/2}{ \cS_{1'4} + \mu^2_{41'23}x^{-1}_{4/1} }
+ 
\frac{-p_{2'}^+/2}{ \cS_{24} + \mu^2_{41'23}x^{-1}_{4/2'} }
\ , \\
R_Z(12;1'2')
&=
\frac{- p_{2}^+/2}{ \cS_{2'4} + \mu^2_{412'3} x^{-1}_{4/2} }
     + \frac{-p_{1'}^+/2}{ \cS_{14} + \mu^2_{412'3} x^{-1}_{4/1'} }
\ .
\end{align}

The other two interaction terms, $U_{t\,\text{eff}}(31;3'1')$
and $U_{t\,\text{eff}}(23;2'3')$, are obtained by cyclic
permutations of $1$, $2$, $3$ and $1'$, $2'$, $3'$ in the
formulas for $U_{t\,\text{eff}}(12;1'2')$.

%%%%%%%%%%%%%%%%%%%%%%%%%%%%%%%%%%%%%%%%%%%%%%%%
 \subsection{ Small-$x$ dynamics }
%%%%%%%%%%%%%%%%%%%%%%%%%%%%%%%%%%%%%%%%%%%%%%%%

The interaction kernel $U_{t\,\rm eff}(12;1'2')$ in 
Eq.~(\ref{QQQeffEVE}) can be written in terms of 
relative momentum variables $x_{1/12}$, $\kappa_{1/12}^\perp$ 
and $x_{1'/12}$, $\kappa_{1'/12}^\perp$ (that is momenta relative
to pair $12$). The resulting expression has the same structure
as in our quarkonium analysis~\cite{QQbarRGPEP}, with the color
factor $2/3$ instead of $4/3$. The gluon mass ansatz in baryons can 
be different from the one in quarkonia, since it is a function 
of gluon momentum relative to the three-quark subsystem 
instead of quark-antiquark subsystem. The small-$x$ singular 
factors in the interaction do not produce divergences for 
the same reason as in quarkonia. The gluon-exchange integral
is finite because we assume that the gluon mass ansatz 
vanishes when $x_5\to 0$. An example of required behavior 
in quarkonia is $\mu^2 \sim x_5^{\delta_\mu} \kappa_5^2$. 
In the notation used for baryons, the same behavior is described
by $\mu^2 \sim x_{4/12}^{\delta_\mu} \kappa_{4/12}^2$
when $x_{4/12}\to 0$. Let us assume that $\mu^2 \sim 
x_4^{\delta_\mu} \kappa_4^2$ in the limit $x_4\to 0$. When 
$x_4$ goes to zero, so does $x_{4/12} = x_4/(x_1+x_2)$,
and $\kappa_{4/12}^\perp \approx \kappa_4^\perp$.
Therefore, $\mu^2 \sim x_4^{\delta_\mu} \kappa_4^2$ 
implies $\mu^2 \sim x_{4/12}^{\delta_\mu} \kappa_{4/12}^2$. 
The same reasoning applies to every pair of the quarks that 
exchange a gluon. Similarly, in the quark self-interaction 
terms, the integration variables are $x_{4/i}$ and $\kappa_{4/i}^\perp$.
In the small-$x_4$ limit, $\kappa_{4/i}^\perp \approx \kappa_4^\perp$ 
and $x_{4/i} \approx x_4 \to 0$. Therefore, mass terms are finite 
when the gluon mass ansatz vanishes properly when $x_4\to0$.

%%%%%%%%%%%%%%%%%%%%%%%%%%%%%%%%%%%%%%%%%%%%%%%%%%%%%%%%%%%%%%%%
 \section{ Effective interactions in the nonrelativistic limit }
 \label{sec:baryonOSC}
%%%%%%%%%%%%%%%%%%%%%%%%%%%%%%%%%%%%%%%%%%%%%%%%%%%%%%%%%%%%%%%%

Given Eq.~(\ref{scales}), the effective Hamiltonian
$H_{{\rm eff}\,t}$ can be approximated by its non-relativistic
(NR) limit. To define this limit, we introduce a set of convenient 
momentum variables, cf. Refs.~\cite{GlazekCondensatesAPP,TrawinskiConfinement},
\beq
Q_3^\perp
\es
\sqrt{\frac{\beta_3(1-\beta_3)}{x_3(1-x_3)}}\ \kappa_{3}^\perp
\rs
\sqrt{\frac{2/9}{x_3(1-x_3)}}\ \kappa_{3}^\perp
\ ,
\label{nrvariables1}
\\
Q_3^z
\es
\sqrt{\frac{\beta_3(1-\beta_3)}{x_3(1-x_3)}}
( x_3 - \beta_3 ) ( m_1 + m_2 + m_3 )
\rs
\frac{\sqrt{2} m ( x_3 - 1/3 )}{\sqrt{x_3(1-x_3)}}
\ ,
\\
K_{12}^\perp
\es
\sqrt{\frac{\beta_1\beta_2(1-x_3)}{x_1 x_2(1-\beta_3)}}
\ \kappa_{1/12}^\perp
\rs
\sqrt{\frac{1-x_3}{6 x_1 x_2}}
\ \kappa_{1/12}^\perp
\ ,
\\
K_{12}^z
\es
\sqrt{\frac{\beta_1\beta_2(1-x_3)}{x_1 x_2(1-\beta_3)}}
\frac{x_1 m_2 - x_2 m_1}{1 - x_3} 
\rs
\sqrt{\frac{1-x_3}{6 x_1 x_2}}
\frac{(x_1 - x_2)m}{1 - x_3}
\ ,
\label{nrvariables4}
\eeq
where $\beta_i = m_i/(m_1 + m_2 +m_3)$. The second equality
in these equations holds only for equal masses. The non-relativistic 
limit is defined as $\vec K/m \to 0$, $\vec Q/m \to 0$. It is valid 
because the relative momentum regions that significantly exceed 
the RGPEP scale $\lambda \ll m$ are suppressed by the exponentially-fast 
vanishing form factors in the interaction vertices of effective particles. 
In the leading NR approximation, the momenta $\vec K_{12}$ and 
$\vec Q_3$ are related to the Jacobi momenta: $\vec K_{12}$ is the 
relative momentum of quark $1$ with respect to $2$ and $\vec Q_3$ 
is the relative momentum of quark $3$ with respect to the pair of 
quarks $1$ and $2$, see Ref.~\cite{GlazekCondensatesAPP} for 
more details. Generically, we denote by $\vec K$ the relative 
momentum of a quark with respect to another quark with which 
it is involved in an interaction term, and we denote by $\vec Q$ the 
relative momentum of a spectator with respect to the pair in 
interaction. We introduce three sets of such relative momentum 
variables, arranged using the cyclic permutation of indices $123$: 
$\vec K_{jk}$ and $\vec Q_i$. In the NR limit, 
\beq
\vec K_{31}
\rs
- \frac{1}{2} \vec K_{12}
+ \frac{3}{4} \vec Q_3
\ ,
\quad
\vec Q_2
\rs
- \vec K_{12}
- \frac{1}{2} \vec Q_3
\ ,
\\
\vec K_{23}
\rs
- \frac{1}{2} \vec K_{12}
- \frac{3}{4} \vec Q_3
\ ,
\quad
\vec Q_1
\rs
\phantom{+} \vec K_{12}
- \frac{1}{2} \vec Q_3
\ .
\eeq
For equal quark masses, we write the baryon mass 
as $M=3m+B$, divide Eq.~(\ref{QQQeffEVE}) by 
$6m$,  take the NR limit and obtain
\begin{align}
\left[
  \frac{ {\vec K_{12}}^{\, 2} }{ 2\mu_{12} }
+ \frac{ {\vec Q_{3}}^{\, 2}  }{ 2\mu_{3(12)} }
- B
+ 3\frac{ \delta m_{1\,t}^2 }{ 2 m }
\right]\,
\psi_t(123)
\nn
+ \sum_{\sigma_{1'}\sigma_{2'}}\int\frac{d^3 K_{12}'}{(2\pi)^3} \, 
  [  f_{t\,12.1'2'} V^{12}_{C,BF}
   + W^{12} ] 
\, \psi_t(1'2'3)
\nn
+ \sum_{\sigma_{2'}\sigma_{3'}}\int\frac{d^3 K_{23}'}{(2\pi)^3} \, 
  [  f_{t\,23.2'3'} V^{23}_{C,BF}
   + W^{23} ] 
\, \psi_t(12'3')
\nn
+ \sum_{\sigma_{3'}\sigma_{1'}}\int\frac{d^3 K_{31}'}{(2\pi)^3} \, 
  [  f_{t\,31.3'1'} V^{31}_{C,BF}
   + W^{31} ] 
\, \psi_t(1'23')
&= 0 \ ,    
\label{eqNR}
\end{align}
where { $V^{ij}_{C,BF} = V_{C,BF}( \vec K_{ij}, \vec K_{ij}'\, )$}
and $W^{ij} = W( \vec K_{ij}-\vec K_{ij}'\, )$ are, respectively, 
the Coulomb term with Breit-Fermi (BF) corrections and the 
additional interaction resulting from the gluon mass ansatz.
$\mu_{12} = m/2$, $\mu_{3(12)} = 2m/3$ are the reduced masses.
Both $V$ and $W$ are similar to the ones in the quarkonium 
case~\cite{QQbarRGPEP}.
\beq
V_{C,BF}(\vec K, \vec K')
\es
- \frac{2}{3} g^2\,
  \frac{1}{\Delta\vec K^2}
  (1+BF)
\ ,
\\
W(\Delta\vec K)
\es
- \frac{2}{3} g^2\left[
    \frac{1}{(\Delta K^z)^2}
  - \frac{1}{\Delta\vec K^2}
  \right]
  \frac{\mu^2}{\mu^2 + \Delta\vec K^2}
  \exp\left[-2 t m^2\frac{\Delta\vec K^4}{(\Delta K^z)^2}\right]
\ ,
\nn
\label{eqforW}
\eeq
where $\Delta\vec K = \vec K - \vec K'$ and the RGPEP form factor is
\beq
f_{t\,ij.i'j'}
\es
\exp\left\{-16 t [\vec K_{ij}^2 - (\vec K_{ij}')^2]^2\right\} \ .
\label{formfactorNR}
\eeq
The mass terms can be written similarly to the interaction 
terms because $\delta m_{1\,t}^2/(2m) = - (2\pi)^{-3}
\int d^3 \tilde K \,  W( \vec{\tilde K})$. Using the Taylor 
expansion for the wave functions under the integrals, {\it e.g.},
\beq
\psi_t(1'2'3)
\es
  \psi_t(123)
+ \Delta K_{12}^l \frac{\pd}{\pd K_{12}^l} \psi_t(123)
\label{taylor}
\\
& + &
\frac{1}{2} \Delta K_{12}^m \Delta K_{12}^n
\frac{\pd^2}{\pd K_{12}^m \pd K_{12}^n} \psi_t(123)
+ \dots \ , \nnn
\eeq
one can see that the first term cancels with a mass term.
Note that in baryons there are three one-gluon-exchange 
terms of Eq.~(\ref{eqforW}) that combine with three quark
self-interaction terms in Eq.~(\ref{eqNR}), while in 
quarkonia there is only one one-gluon-exchange term that 
combines with two quark self-interaction terms. The first 
term in Eq.~(\ref{taylor}) and the self-interaction terms 
combine in baryons as in quarkonia in Ref.~\cite{QQbarRGPEP}, 
because the color factors for the gluon-exchange terms in 
baryons are twice smaller than in quarkonia, while the color 
factors for quark self-interactions are the same in both 
systems. The second term is linear in momentum $\Delta\vec K_{12}$ 
and gives zero after integration. The first non-vanishing 
term is the third one, quadratic in $\Delta\vec K_{12}$. 
This term provides a harmonic oscillator potential. Only 
terms with $m=n$ are non-zero, and
\begin{align}
\int\frac{d^3\! K_{12}'}{(2\pi)^3}  W^{12}
\, [ \psi(1'2'3) - \psi(123) ]
\approx
- w^n \frac{\pd^2 \psi(123)}{\pd (K_{12}^n)^2}
\, .
\label{wfexpansion}
\end{align}
As in quarkonia, we assume that the ansatz $\mu^2$ 
dominates $\Delta\vec K^2$ in the relevant integration 
range. In this case, in Eq.~(\ref{eqforW}), $\mu^2/(\mu^2 +
\Delta\vec K^2)\approx 1$, which further leads to the 
conclusion that $w^n$ for $n=1,2,3$, corresponding to 
different directions in space, are the same. Thus, the 
effective oscillator interaction respects rotational symmetry 
in the Jacobi variables. However, there are only two 
independent relative momenta for three quarks. We 
distinguish one pair of quarks, {\it e.g.}, $12$, and rewrite 
the oscillators in terms of $\vec K_{12}$ and $\vec Q_3$,
\begin{align}
  \left(\frac{\pd}{\pd\vec K_{12}}\right)^2
+ \left(\frac{\pd}{\pd\vec K_{23}}\right)^2
+ \left(\frac{\pd}{\pd\vec K_{31}}\right)^2
\rs
  \frac{3}{2}
  \left(\frac{\pd}{\pd\vec K_{12}}\right)^2
+ 2 \left(\frac{\pd}{\pd\vec Q_{3}}\right)^2
\ .
\label{HOsrewrittenQQQ}
\end{align}
Thus, we obtain an oscillator force between quarks
$1$ and $2$ and an oscillator force between quark 
$3$ and the pair $12$. Their strengths are in ratio
$3/2$ : $2$. Since the ratio of corresponding reduced 
masses $\mu_{3(12)}$ and $\mu_{12}$ is $4/3$,
the frequencies of these oscillators are the same 
and equal
\beq
\omega_\text{baryon}
\es
\frac{\sqrt{3}}{2} \sqrt{\frac{\alpha}{18\sqrt{2\pi}}}
\
\frac{\lambda^3}{m^2}
\ .
\label{obequal}
\eeq
This expression differs from the result for quarkonia
by a factor $\sqrt{3}/2$, rendering $\omega_\text{baryon}^2/
\omega_\text{meson}^2=3/4$, assuming that $m$,
$\lambda$ and $\alpha$ are the same for mesons 
and baryons built from one flavor of heavy quarks.
This result is very close to the ratio 5/8 suggested 
by models that employ the concept of gluon condensate
in vacuum~\cite{GlazekSchadenCondensate} or only
inside hadrons~\cite{GlazekCondensatesAPP}. 

The oscillator interaction may appear to be in contradiction 
with the linear confinement picture in QCD. However, the 
eigenvalues of the FF Hamiltonian are the baryon masses 
squared, in distinction from the instant form (IF) Hamiltonian 
eigenvalues that are the baryon energies, reducing to the 
baryon masses only for bound states at rest.  At large 
distances between quarks, the quadratic potential in the FF 
corresponds to the linear potential in the IF of Hamiltonian 
dynamics~\cite{TrawinskiConfinement}. 

%%%%%%%%%%%%%%%%%%
 \section{ Two flavors of heavy quarks }
 \label{sec:twoflavors}
%%%%%%%%%%%%%%%%%%

Several new elements appear when one of the three quarks,
say quark $3$, is of different flavor than the other two.
Besides smaller particle-exchange symmetry and the fact 
that $\beta_1\neq\beta_3\neq1/3$, a new feature emerges 
that the NR effective quark masses are modified. 

The reason for NR mass modification is that, when we deal 
with two different flavors of quarks, the constant term that 
cancels completely in Eq.~(\ref{wfexpansion}) for the same
flavor no longer does so for different flavors. A finite function 
of $x$ and $\kappa^\perp$ is left and it multiplies $\psi(1,2,3)$. 
This effect is small, but in principle ought to be considered. 
The correction shifts the minimal invariant mass squared 
value around which the NR approximation is obtained. Namely, the 
optimal values of $\beta_i$ around which one expands are 
slightly altered, {\it cf.} Eqs.~(\ref{nrvariables1}) to (\ref{nrvariables4}). 
The shifts spoil the rotational symmetry of the second-order 
Coulomb and harmonic oscillator potentials. For $b$ and $c$ 
quarks, the deviation from spherical symmetry appears to be 
on the order of a few percent. It depends on the gluon mass. 

This effect is certainly going to change in calculations 
of higher order than second, because it depends on the gluon 
mass ansatz and the ansatz will be replaced by theory. Since 
this effect is relatively small, we neglect it in what follows. 
Apart from the neglected effect, the Coulomb interactions 
between quarks are not altered when flavors differ. 

The harmonic oscillator forces between pairs of quarks 
depend on the quark masses. Instead of Eq.~(\ref{HOsrewrittenQQQ}),
in which a common coefficient $w$ is omitted, one obtains 
\begin{align}
&
  w_{12}\left(\frac{\pd}{\pd\vec K_{12}}\right)^2
+ w_{23}\left(\frac{\pd}{\pd\vec K_{23}}\right)^2
+ w_{31}\left(\frac{\pd}{\pd\vec K_{31}}\right)^2
\nn
&
=
\left( w_{12} + \frac{1}{2} w_{23} \right)
  \left(\frac{\pd}{\pd\vec K_{12}}\right)^2
+ 2 w_{23} \left(\frac{\pd}{\pd\vec Q_{3}}\right)^2
\ ,
\end{align}
where 
\beq
w_{ij}
\es
\frac{\alpha \lambda^3 }{18\sqrt\pi}\left(\frac{\lambda^2}{m_i^2+m_j^2}\right)^{3/2}
\ ,
\eeq
and $w_{31}=w_{23}$. Using the reduced masses
$\mu_{12}$ and $\mu_{3(12)}$, one can write the 
frequencies squared for the Jacobi oscillation modes 
$12$ and $3(12)$ as
\begin{align}
\omega_{12}^2
&=
\frac{1}{m_1}
\frac{\alpha \lambda^3}{18\sqrt\pi}
\left[
  \left(\frac{\lambda^2}{2m_1^2}\right)^{3/2}
+ \frac{1}{2}\left(\frac{\lambda^2}{m_1^2+m_3^2}\right)^{3/2}
\right] ,
\label{diquarkomega}
\\
\omega_{3(12)}^2
&=
\frac{2m_1+m_3}{2m_1 m_3}
\frac{\alpha \lambda^3 }{18\sqrt\pi}
\left(\frac{\lambda^2}{m_1^2+m_3^2}\right)^{3/2} .
\label{omega3}
\end{align}

The frequencies depend on the quark masses and the RGPEP 
scale parameter $\lambda$. However, one can expect that there 
exists a window of values of $\lambda$, in which eigenvalues of 
the approximate effective Hamiltonians $H_{{\rm eff}\,t}$ are 
close to the eigenvalues of the exact renormalized Hamiltonian 
$H_t$ (which do not depend on $\lambda$)~\cite{GlazekWilson1998,GlazekMlynikAF}. The window should 
broaden in higher order calculations due to new interactions 
that possibly appear in $H_{{\rm eff}\,t}$ and due to running 
of effective masses and couplings. The hope is that, similarly 
as in matrix models, the $H_{\rm{eff} \, t}$ that is calculated 
using the low-order weak-coupling perturbative expansion for 
Hamiltonian operators, grasps the main features of bound states 
in QCD despite the growth of the coupling constant when $\lambda$ 
is lowered.

The effective eigenvalue equation for heavy baryons 
in QCD of two heavy flavors, implied by our gluon 
mass hypothesis, is 
\beq
&& \left[
  \frac{K_{12}^2}{2 \mu_{12}}
+ \frac{Q_{3}^2}{2 \mu_{3(12)}}
- \frac{\mu_{12} \omega_{12}^2 \Delta_K^2}{2}
- \frac{\mu_{3(12)} \omega_{3(12)}^2 \Delta_Q^2}{2}
\right]
\psi_t(\vec K_{12}, \vec Q_3)
\np
\int\frac{d^3q}{(2\pi)^3} V_C(q) \
  \psi_t\left(\vec K_{12}-\vec q, \vec Q_3 \right)
% \np
+
 \int\frac{d^3q}{(2\pi)^3} V_C(q) \
  \psi_t\left(\vec K_{12}+\frac{1}{2}\vec q, \vec Q_3+\vec q \right)
\np 
\int\frac{d^3q}{(2\pi)^3} V_C(q) \
  \psi_t\left(\vec K_{12}+\frac{1}{2}\vec q, \vec Q_3-\vec q \right) 
% \nn
\rs
E\ \psi_t(\vec K_{12}, \vec Q_3)
\ ,
\label{EVE}
\eeq
where $\Delta$ denotes Laplacian, reduced masses are  
$ \mu_{12} = m_1/2$, $ \mu_{3(12)} = 2 m_1 m_3 / (2m_1+m_3)$
and
\beq
V_C(q)
\es
-\frac{2}{3} \frac{g^2}{q^2} \ .
\eeq
The baryon mass eigenvalue is obtained from the eigenvalue $E$,
\beq
M
\es
(2m_1+m_3)\sqrt{1+\frac{2 E}{2m_1+m_3}}
\ .
\label{massbaryon}
\eeq
We omitted the BF spin-dependent terms and the RGPEP 
form factors whose numerical inclusion requires the 
fourth-order RGPEP calculation. So, Eq.~(\ref{EVE}) 
only accounts for interactions of order $\alpha$. The 
associated quarkonium eigenvalue equation is~\cite{QQbarRGPEP}
\beq
\left[
  \frac{K_{12}^2}{2 \mu_{12}}
- \frac{\mu_{12} \omega_{12}^2 \Delta_K^2}{2}
\right] \psi_t(\vec K_{12})
+ 2 \int\frac{d^3q}{(2\pi)^3} V_C(q) \,
  \psi_t(\vec K_{12}-\vec q)
\rs
E \ \psi_t(\vec K_{12})
\ ,
\nn
\label{EVEm}
\eeq
\beq
M \rs (m_1 + m_2) \sqrt{ 1 + \frac{2E}{m_1 + m_2}} \ ,
\eeq
where $\omega_{12}^2$ is given in Eq.~(\ref{omegabc}).

%%%%%%%%%%%%%%%%%%%%%%
\section{ Sketch of triply heavy baryon spectra }
\label{sec:sketch}
%%%%%%%%%%%%%%%%%%%%%%

In this paper, we focus on qualitative features of our method, 
and test its capability to describe baryons. Therefore, we only sketch 
the spectrum of heavy baryons that follows from QCD of quarks 
$b$ and $c$ including our pilot simplifications, the latter 
being gradually removable increasing the order of weak coupling
expansion for $H_{{\rm eff} \, t}$ and number of effective Fock 
components in the eigenvalue problem of $H_{{\rm eff} \, t}$. 

In order to solve the baryon bound-state problem one needs to 
fix $\alpha$, $m_c$ and $m_b$. To estimate these quantities in the
RGPEP scheme, we use data for heavy quarkonia. The first issue 
one needs to deal with is the strong dependence of oscillator 
frequencies on $\lambda$. If our calculations of $H_t$ and its 
eigenvalues were exact, the observables we obtain would be independent 
of $\lambda$, which hence could be chosen arbitrarily. Since we 
solve the RGPEP equation only up to order $\alpha$ and we introduce 
a gluon mass ansatz to reduce the eigenvalue problem to the hadron 
dominant Fock component, the effective dynamics we obtain may 
provide a reasonable approximation only in a certain window of 
values of $\lambda$ (see particularly Fig.~4 in Ref.~\cite{GlazekWilson1998} 
and Fig.~4 in Ref.~\cite{GlazekMlynikAF}).

We discuss the approximate hadron spectra that our method produces 
from heavy-quark QCD using the assumption that $\lambda \sim 
\sqrt{\alpha} \  m_Q$, where $m_Q$ is a suitable quark mass 
parameter (we introduce its definition in Sec.~\ref{sec:params}). 
When $\alpha$ is sufficiently small, this assumption fulfills 
constraints of Eq.~(\ref{scales}) and, moreover, for quark--antiquark 
system it ensures that $\lambda \gg k_B \sim \alpha \,\mu$, where 
$k_B$ is the strong Bohr momentum and $\mu$ is the quark reduced 
mass, which in turn ensures that RGPEP form factors do not influence 
significantly the eigenvalues of $H_{{\rm eff}\,t}$. For example, 
if $\lambda$ were equal $\alpha \mu \ll \sqrt{\alpha}\mu$, then 
the form factor of Eq.~(\ref{formfactorNR}) for $K = 0$ and $K' = 
\alpha\mu$ would be $f = e^{-16}$, which is practically zero, no 
matter how small the coupling constant is, and the form factor
would play a significant role in the eigenvalue problem. On the 
other hand, if $\lambda = \sqrt{\alpha}\mu$, then, for the same 
$K$ and $K'$ as before, $f = e^{-16\alpha^2}$, which is practically 
one for sufficiently small $\alpha$, and the form factor is invisible 
in the first approximation. Such assumption is well justified in QED, 
where the Schr\"odinger equation with simple local Coulomb potential 
gives very good first approximation to the Hydrogen spectrum. 
Please note, however, that the form factors are necessary in 
higher order calculations, because they make interactions, which 
otherwise would be singular, like spin-spin interactions, finite, 
and actually small in comparison to the leading binding effects 
that we describe below.

Note also that keeping $\lambda$ proportional to the square root 
of $\alpha$ secures proportionality of the resulting hadron 
binding energies to $\alpha^2$~\cite{Wilsonetal}, which
resembles analogous scaling in QED. This scaling is maintained 
with our oscillator terms because their frequencies emerge 
proportional to $\alpha^2$. Knowing that the observed low-mass 
quarkonium spectra can be characterized as intermediate between 
the Coulomb and oscillator spectra~\cite{Eichten:1978tg}, 
we expect that the harmonic oscillator frequencies obtained 
from QCD may be comparable in size with the strong-interaction 
Rydberg-like constant $R = \mu (4\alpha/3)^2/2$. 

Finally, a comment is in order regarding our use of perturbation 
theory for calculating effective Hamiltonian while the coupling 
constant is to be extrapolated from an infinitesimal to a finite 
value,  Formally, the whole calculation is valid only in the limit 
of infinitesimal coupling constant (or $\Lambda_{\rm QCD} \to 0$), 
because only then one can consider perturbative terms, {\it e.g.}, 
the divergent mass counterterm, as small perturbations. Therefore, 
we stress that we assume that for the values of $\alpha$ between 
about one quarter and one half (depending on the system under 
consideration) the functional form of effective Hamiltonians as 
a function of $\alpha$, does not change 
significantly~\cite{GlazekWilson1998,GlazekMlynikAF}. 

The numbers we obtain are listed including four or even five 
significant digits only because the data we approximately 
reproduce~\cite{PDG2016} provide that many digits. We ignore 
data error bars and use our analytic expressions. The Coulomb 
effects are estimated in first-order perturbation theory around 
the relevant oscillator solutions. Only diagonal matrix elements 
need to be considered because inclusion of non-diagonal matrix 
elements produces relatively small effects that do not change the main
features of lowest-mass heavy-baryon spectrum that we sketch. For example, the ground states of $ccc$ and $bbb$ 
shift by about $31$~MeV and $47$~MeV respectively when instead 
of first order perturbation theory one diagonalizes Hamiltonian 
matrix in the basis of harmonic oscillator eigenstates with 
excitation energy up to $4\omega_\text{baryon}$. These corrections
are small in comparison with the expected effects of spin dependent 
interactions, which we neglect (but we do include effects due 
to the Pauli exclusion principle for fermions). Therefore, given 
the simplifications we have made in the pilot application of our 
method to solving heavy-flavor QCD, we provide a sketch of the 
low-mass hadron spectra obtained from first-order perturbation 
theory around the oscillator spectrum implied by the assumption 
that gluons develop a mass. 

%%%%%%%%%%%%%%%%%%%%%%%%%%%%%%%%%%%%%%%%%%%%%%%%
 \subsection{ Adjustment of parameters $\alpha$, $m_b$, $m_c$ and $\lambda$ }
 \label{sec:params}
%%%%%%%%%%%%%%%%%%%%%%%%%%%%%%%%%%%%%%%%%%%%%%%%

The coupling constant dependence on $\lambda$ is 
set to the well-known approximate function, {\it cf.} 
Ref.~\cite{AF},
\beq
\alpha \es  \left[ \beta_0 \log (\lambda^2/\Lambda^2_{\rm QCD})  \right]^{-1} \ , 
\label{alphaQCD}
\eeq
where $\beta_0 = (33 - 2 n_f)/(12 \pi)$ and $n_f=2$, valid 
in QCD of two heavy flavors $b$ and $c$, ignoring $u$, $d$, 
$s$ and $t$. Demand that $\alpha = 0.1181$ for $\lambda
= M_Z = 91.1876$ GeV, would enforce the RGPEP value of 
$\Lambda_{\rm QCD} = 371$ MeV and we use this value. 
The resulting spectra do not change significantly when we 
change $n_f$ in the range from $2$ to $5$.

The quark masses are assumed to be independent of $\lambda$
because their dependence is not known yet in the RGPEP. 
Confinement poses a conceptual difficulty concerning the 
definition of quark mass~\cite{PDG2016}. Quantitative 
estimates of quark masses would need the RGPEP calculation
to at least fourth order while we consider only second. 
Formulas for running masses of quarks in other approaches, 
like in Eq.~(9.6) in~\cite{PDG2016}, do not concern mass 
terms in the FF Hamiltonian $H_t$. At the current level of 
crude approximation and not knowing the masses precisely, 
we assume that $m_b$ and $m_c$ can be treated as constants 
in the range of values of $\lambda$ that we use in fitting 
data.

In the case of quarkonia, we set
\beq
\lambda_{Q \bar Q} \es \sqrt{\alpha}
\left( a \,  \bar m_{Q\bar Q} + b \right) \ ,
\label{lambdaQQ}
\eeq
where $\bar m_{Q \bar Q}$ is the average mass of quark 
and antiquark that form a heavy meson, such as $J/\psi$, 
$\Upsilon$ or $B_c^+$. The quark masses and unknown 
values of $a$ and $b$ are fitted
to the spectra of heavy quarkonia. Separate fits for
a set of $b\bar b$ states and a set of $c \bar c$ states
give us most suitable $\lambda_{b \bar b}$ and $\lambda_{c \bar c}$,
and quark masses $m_b$ and $m_c$. This set of numbers
allows us to fix values of $a$ and $b$
in the linear formula of Eq.~(\ref{lambdaQQ}). With $a$ 
and $b$ fixed, we test Eq.~(\ref{lambdaQQ}) by 
comparing our theoretical spectrum of $B_c$ particles, 
computed for $\lambda_{bc}$ given by Eq.~(\ref{lambdaQQ}), 
with experimental data. The agreement is satisfactory, 
{\it cf.}~Sec.~\ref{sec:qqmasses}. The adjustment of constants 
$a$ and $b$ reflects the current lack of 
knowledge of the values of $\lambda$ at which one can most 
accurately approximate different hadron eigenvalue problems 
using merely their lowest Fock components and gluon mass 
ansatz. Details of our fits of two quark masses, $m_c$ and 
$m_b$ at most suitable values of $\lambda_{b \bar b}$ and 
$\lambda_{c \bar c}$, are described in \ref{app:fits}.

In the case of baryons, we set
\beq
\lambda_{3Q} \es \sqrt{\alpha}
\left( a \,  \bar m_{3Q} + b \right) \ ,
\label{lambda3Q}
\eeq
where $\bar m_{3Q}$ is the average mass of the three quarks 
that form a lowest Fock component of a baryon at scale
$\lambda_{3Q}$. The linear formula secures that $\lambda_{bbb} = 
\lambda_{b\bar b}$ and $\lambda_{ccc} = \lambda_{c \bar c}$ 
and it means that no exotic changes occur in between. With 
the linear interpolation, for which no alternative has been 
identified, it turns out that our estimates for $bbb$ and 
$ccc$ spectra resemble results of other approaches, see below. 

Our estimates are quite crude. We ask two questions. One is
if the oscillator terms that follow from the assumption of
gluon mass are capable of providing a reasonable first 
approximation to heavy hadrons. Provided that in the case of 
heavy quarkonia the answer is yes, the other question is what 
character of the heavy baryons spectrum one expects using the 
assumption that effective gluons develop a mass. To address 
these qualitative questions, we ignore the BF spin-dependent 
terms and we estimate strong-Coulomb effects by evaluating 
expectation values of the corresponding interaction terms in 
the oscillator eigenstates. Details of unperturbed baryon 
wave functions are described in \ref{app:wavefunctions}. Comparison 
with other approaches, including lattice estimates, suggests 
that our extremely simple oscillator picture and thus possibly 
also the gluon mass hypothesis, appear reasonable. Reliable 
estimates of better accuracy require fourth-order solution to 
the RGPEP Eq.~(\ref{RGPEP}).

%%%%%%%%%%%%%%%%%
 \subsection{ Masses of quarkonia }
 \label{sec:qqmasses}
%%%%%%%%%%%%%%%%%

\begin{figure*}
\centering
\includegraphics{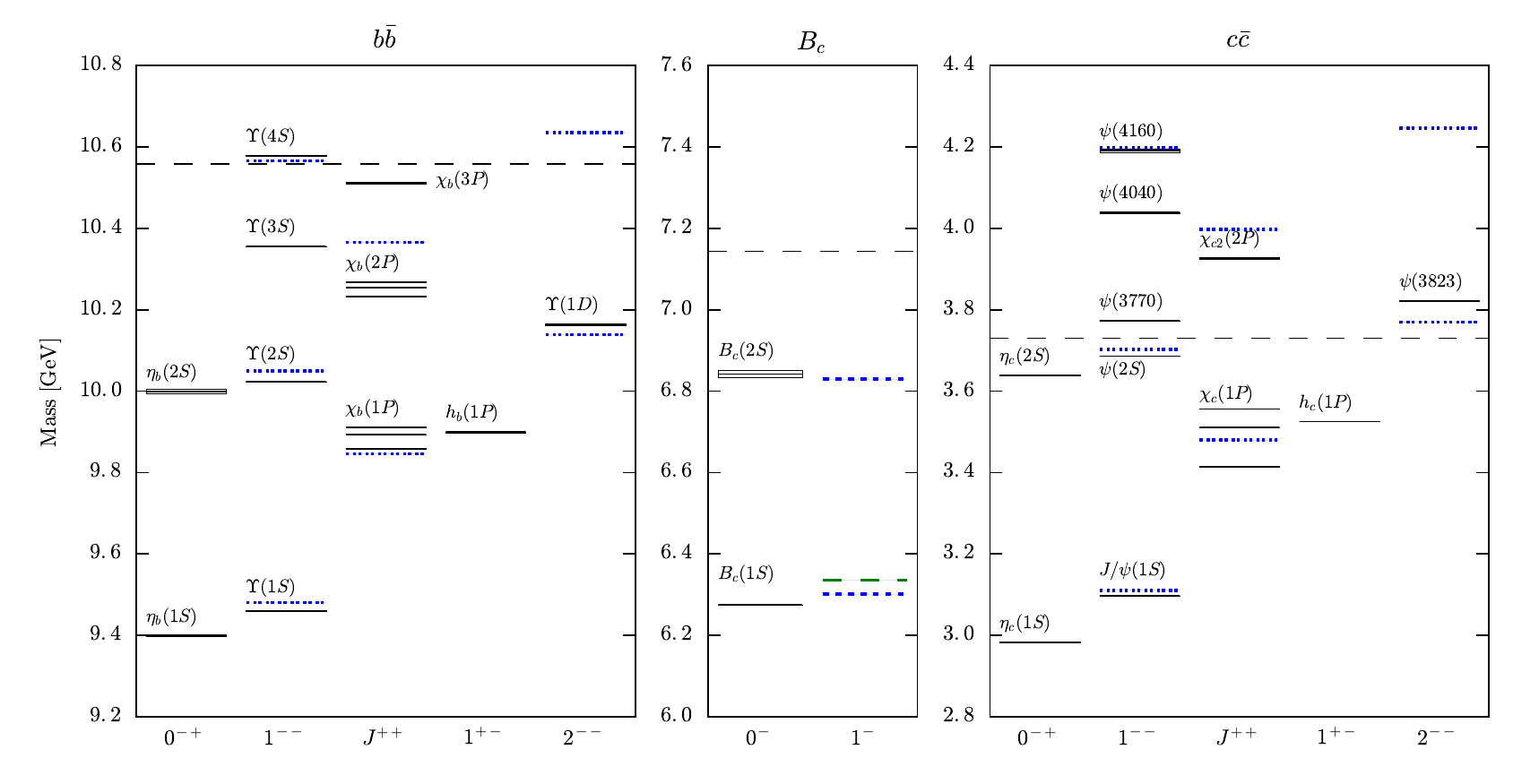}
\caption{ Results of our fit to $b \bar b$ and $c \bar c$ low-mass
states, are shown by dotted blue lines on the background of data for 
heavy quarkonia and $B_c$ mesons. The dashed green line 
represents an average of various predictions for $B_c^*$ 
mass~\cite{Gomez-Rocha:2016cji}. Dashed lines across a panel 
represent thresholds. We include the states $3S$ that appear 
at the level of observed states that are interpreted as $4S$.
Such highly excited states may contain an important component
with effective gluons, not properly accounted for by the perturbative
calculation in our pilot study. } 
\label{fig:qqspectra}
\end{figure*}

Details of fits of quark masses and scale parameter 
to quarkonium data are described in \ref{app:fits}. 
Most accurate fit to masses of $\Upsilon(1S)$,
$\Upsilon(2S)$ and $\chi_{b1}(1P)$, is obtained for
\beq
m_b
\es
4698\text{ MeV}
\quad \text{and} \quad
\lambda_{b\bar b}
\rs
4258\text{ MeV}
\ .
\eeq
These values are associated with $\alpha(\lambda_{b\bar b}) = 0.2664$
and $\omega_{b\bar b} = 268.8$ MeV. The resulting bottomonium 
masses are shown in the left panel of Fig.~\ref{fig:qqspectra}.
To most accurately describe masses of $J/\psi$, $\psi(2S)$ and 
$\chi_{c1}(1P)$, one needs 
\beq
m_c
\es
1460\text{ MeV}
\quad \text{and} \quad
\lambda_{c\bar c}
\rs
1944\text{ MeV}
\ ,
\eeq
and these values are associated with 
$\alpha(\lambda_{c\bar c}) = 0.3926$
and
$\omega_{c\bar c} = 321.6$ MeV.
The resulting charmonium masses are illustrated in
the right panel of Fig.~\ref{fig:qqspectra}.

Values of $\lambda_{b \bar b}$, $\lambda_{c \bar c}$, and
$m_b$, $m_c$ fix
\beq
a \es 1.589 \ ,
\label{aQQbar}
\\
b \es 783\text{ MeV} \ .
\label{bQQbar}
\eeq
These coefficients imply, according to Eq.~(\ref{lambdaQQ}),
\beq
\lambda_{bc} \es 3134\text{ MeV} \ .
\label{lambdabc}
\eeq
The middle panel of Fig.~\ref{fig:qqspectra} shows the comparison
of experimental masses of $B_c$ and $B_c(2S)$ and an average of
different predictions for a mass of $B_c^*$ with our theoretical
levels. Note that because we fit the masses of spin-one quarkonia 
while we neglect spin-dependent interactions, we present our 
mass estimates as for $1^-$. Because there are no experimental 
data to compare for spin-one $B_c^*$, we provide an average of 
various theoretical predictions~\cite{Gomez-Rocha:2016cji}. The 
agreement is satisfactory, given that we do not expect our estimates
to be precise. Equation~(\ref{lambdabc}) gives
\beq
\alpha(\lambda_{bc})
\es
0.3047
\ ,
\\
\omega_{bc}
\es
261.1\text{ MeV}
\ .
\eeq

%%%%%%%%%%%%%%%%%%%%%%%%
\subsection{ Estimates of masses of heavy baryons }
%%%%%%%%%%%%%%%%%%%%%%%%

\begin{figure}[ht!]
\centering
\includegraphics{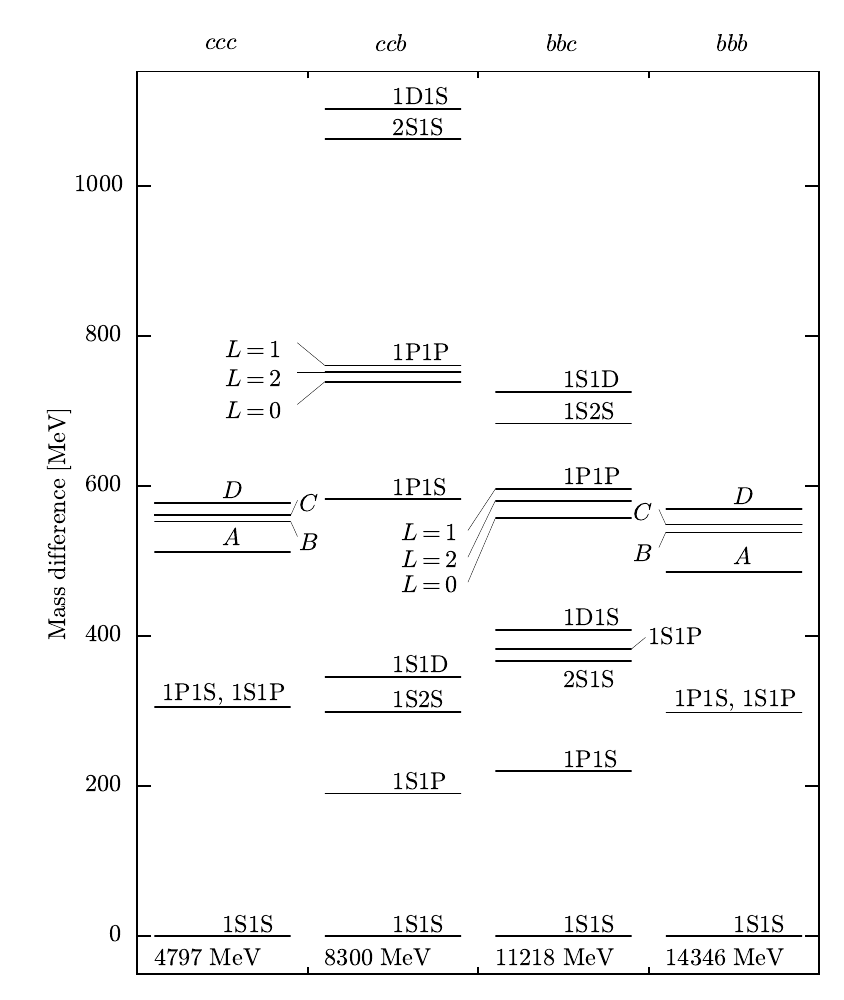}
\caption{ Qualitative picture of triply heavy baryon mass 
spectrum implied by the second-order RGPEP in heavy-flavor 
QCD and our gluon mass ansatz. The figure shows excitations
above the ground states $1S1S$, whose absolute masses are
written at the bottom of each column. The $ccb$ spectrum
displays extraordinarily large mass excitation for states $1P1S$, 
$2S1S$ and $1D1S$ in $ccb$. Such high excitations are 
associated with formation of $cc$-diquarks, bound by a 
harmonic force that is strong because the charmed quarks 
are much lighter than the bottom quarks. Much less pronounced 
splittings appear in the $bbc$ baryons.  See the text for 
further discussion. }
\label{fig:QQQ}
\end{figure}

The fit to quarkonia described in Sec.~\ref{sec:qqmasses} establishes 
optimal values of $\lambda$ for all baryons. Values of the coupling
constant are obtained from Eq.~(\ref{alphaQCD}). The optimal values 
we obtain for these parameters are listed in \ref{app:predictions}. 
The resulting masses of heavy baryons are shown in Fig.~\ref{fig:QQQ}. 
Labels of states describe internal orbital motion of quarks, where the 
first part of a label corresponds to the motion of quark $1$ with respect 
to quark $2$ and the second part corresponds to the motion of quark 
$3$ with respect to the pair of quarks $1$ and $2$. For example, in 
the state $1P1S$, the pair $12$ is in a $p$-wave without radial excitation, 
while the quark $3$ in its motion with respect to the pair $12$ is in an 
$s$-wave state without radial excitation. In $1S2S$, both $1$ with 
respect to $2$ and $3$ with respect to $12$ are in an $s$-wave state 
but the latter is radially excited. States $A$, $B$, $C$ and $D$ in 
$ccc$ and $bbb$ correspond to the second excitation of harmonic 
oscillator with excitation energy $2\omega$ above the ground state 
($\omega\equiv\omega_\text{baryon}$). These states have spin-momentum 
wave functions that are symmetrized in a way due for fermions in 
colorless states. Details of the harmonic oscillator basis wave 
functions are described in \ref{app:wavefunctions}. Analytical 
formulas for masses of baryons are given in \ref{app:masses}.

The values of masses we obtain for $bbb$ and $ccc$ bar\-yons agree 
well with model calculations~\cite{Bjorken:1985ei,Tsuge:1985ei,SilvestreBrac:1996bg,Jia:2006gw,Martynenko:2007je,Roberts:2007ni,Vijande:2015faa}
including quark-diquark~\cite{Giannuzzi:2009gh}
and hypercentral approximations~\cite{Ghalenovi:2014swa,Shah:2017jkr},
bag models~\cite{Ponce:1978gk,Hasenfratz:1980ka,Bernotas:2008bu},
Regge phenomenology~\cite{Wei:2015gsa,Wei:2016jyk},
sum rules~\cite{Zhang:2009re,Wang:2011ae,Aliev:2012tt,Aliev:2014lxa},
pNRQCD~\cite{LlanesEstrada:2011kc},
Dyson-Schwinger approach~\cite{SanchisAlepuz:2011aa,Qin:2018dqp}
and lattice studies~\cite{Brown:2014ena,Meinel:2012qz,Briceno:2012wt,Padmanath:2013zfa,Namekawa:2013vu,Alexandrou:2014sha,Can:2015exa},
where comparison is available.
As an example of comparison, we note that the ground state
of $ccc$ is assigned masses from 4733~MeV to 4796~MeV,
by different lattice calculations, with an average of 4768~MeV. 
Our result is 4797~MeV, differing by 29~MeV, or 0.6~\% from 
the average. For $bbb$, the average of two lattice results we
have identified is 14369~MeV, and our result is 14346~MeV, 
which is 23~MeV difference, or 0.2~\%. These comparisons 
refer to Table I in Ref.~\cite{Shah:2017jkr} that summarizes 
results of calculations of masses of $\Omega_{ccc}$ and 
$\Omega_{bbb}$ reported in twenty different articles.
Ground state of $bbc$ is also very close to the lattice
result~\cite{Brown:2014ena}. In contrast, our $ccb$ differs 
by about 300~MeV from the lattice. We comment on this 
feature below. Comparison with lattice calculations reported in 
Ref.~\cite{Meinel:2012qz} shows that our splittings in $bbb$ 
differ only by about 10~\%. In case of $ccc$~\cite{Padmanath:2013zfa},
the difference of splittings does not exceed 20~\%. This degree 
of agreement is surprising in view of the complexity of lattice 
calculations in comparison with the simplicity of our effective 
Hamiltonian calculation.

The prominent feature visible in Fig.~\ref{fig:QQQ}
is the extraordinary magnitude of splittings in the
$ccb$ baryons. It is a consequence of large oscillator
frequency in $cc$ subsystem in Eq.~(\ref{diquarkomega}),
due to large ratio of $\lambda_{ccb}/m_c$, in which 
the scale parameter $\lambda_{ccb}$ is large in 
comparison with $m_c$ due to $m_b$. This separation 
of scales may make precise calculations of masses of 
excited $ccb$ baryons difficult. Since the harmonic 
excitation is so high, it is likely that components with 
gluons of mass on the order of 1 GeV have to be included 
in a nonperturbative way.  

\begin{figure}[ht!]
\centering
\includegraphics{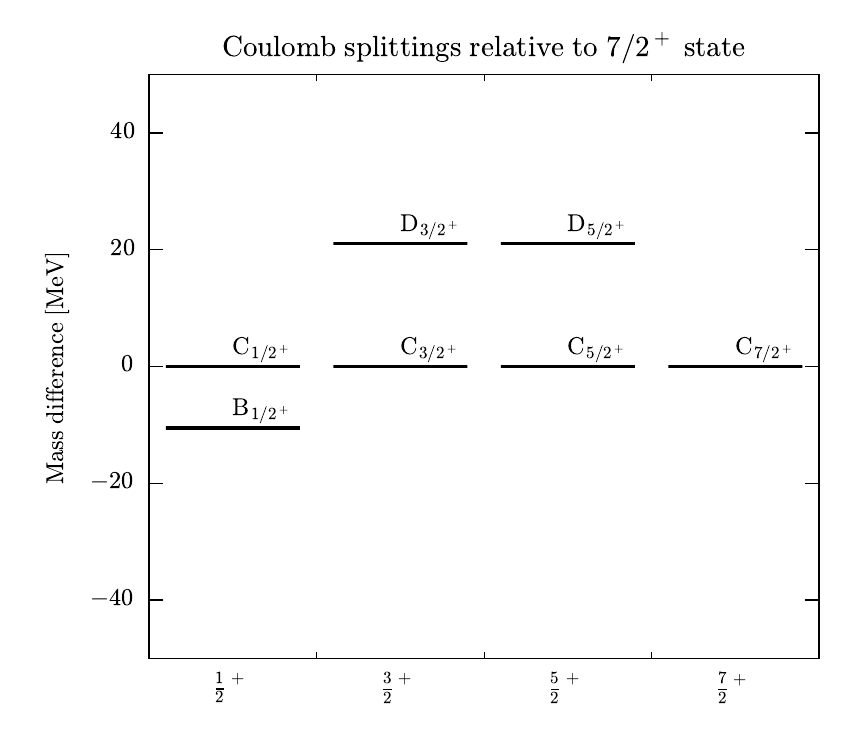}
\caption{ Expanded view of $B$, $C$ and $D$ states of $bbb$
in Fig.~\ref{fig:QQQ}. The harmonic oscillator and Coulomb 
potential are sufficient to qualitatively reproduce the pattern 
of splittings presented in Fig.~11 in Ref.~\cite{Meinel:2012qz}. 
However, quantitatively our pilot-study splittings are about twice 
smaller. Precise agreement is not expected because our 
approximation is very crude and does not include spin interactions. 
Hence, it cannot match splittings between the $C$ states and 
between the $D$ states obtained in other calculations that 
include spin effects. }
\label{fig:bbbsplittings}
\end{figure}

The surprising feature that the very crude, first approximation 
based on the RGPEP, with no free parameters left after adjusting 
quark masses and scale to $b \bar b$ and $c \bar c$ data, 
produces in an elementary analytic way similar splittings to 
the ones resulting from advanced calculations, is further 
illustrated in Fig.~\ref{fig:bbbsplittings}. It presents splittings 
in a second band of harmonic oscillator caused by Coulomb 
interactions. Splittings $m_D-m_C$, $m_C-m_B$, $m_B-m_A$ 
are in relation 2:1:5, which is the general result in the first order 
of perturbation theory for harmonic oscillator perturbed by any 
potential~\cite{Isgur:1978wd}. 

Interestingly, analogous lattice QCD splittings with spin 
dependent interactions turned off~\cite{Meinel:2012qz}, also
appear in the ratios 2:1:5. These results suggest that the 
RGPEP constituent picture with a gluon mass ansatz may be 
grasping the physics of lowest-mass heavy baryons. Since 
experimentally triply heavy baryons are difficult to produce 
and detect~\cite{Olsen:2017bmm}, their theoretical 
understanding using standard techniques is weakly motivated 
and hence also limited~\cite{Brambilla:2010cs,Brambilla:2014jmp}.
Therefore, the ease with which our method yields results for 
heavy baryons in agreement with complex approaches suggests 
that application of the RGPEP in fourth order and including 
components with one or more effective gluons in the eigenvalue 
problem beyond perturbation theory, are worth attempting.

%%%%%%%%%%%
 \section{ Conclusion }
 \label{sec:conclusion}
%%%%%%%%%%%

The effective Hamiltonians we finesse for heavy quarkonia 
and baryons from QCD of charm and beauty quarks using
our gluon mass ansatz, lead to the baryon mass spectra in 
the ball park of expectations from other approaches to physics 
of $ccc$ and $bbb$ systems. In addition, the Hamiltonians 
suggest that quarks $c$ form tight diquarks in $ccb$ baryons.
Diquarks are less likely in $bbc$ baryons. Other approaches do 
not foresee tight diquarks in $ccb$ baryons. This feature may
thus distinguish a physically proper approach in future. However, 
such tight diquarks are hard to excite and mass splittings due 
their excitation are comparable or even exceed values of the 
gluon mass one may expect in theory. In that case, the highly
excited baryon component with a heavy gluon may be large 
and our approximation to the three-quark component as 
dominant may be invalid. Calculations that treat the highly 
excited baryons as having significant components with one 
heavy effective gluon may yield smaller masses than our 
approximation based on the dominance of the three-quark 
component. If it were the case, the RGPEP approach would
still apply, but in the domain of hadron physics in which gluons
appear as constituents in competition with quarks for probability 
of appearance.

Taking into account that the method of RGPEP that we use is 
invariant under boosts and that it is {\it a priori} capable of 
providing a relativistic theory of hadrons in terms of a limited 
number of their effective constituents with suitably adjusted 
size, an extension of the RGPEP calculation to fourth order 
appears worth undertaking. It is certainly needed for verifying 
if the gluon mass ansatz we introduced provides an adequate 
representation of dynamics of gluons in the presence of heavy 
color sources. Fourth-order Hamiltonian is also needed for 
control on the spin splittings and rotational symmetry.

The ratio $\sqrt{8/6}$ of harmonic oscillator frequencies in heavy 
quarkonia and triply heavy baryons is close to the ratio $\sqrt{8/5}$
obtained for $u$ and $d$ constituent quarks in models using the 
concept of gluon condensate. If this is not accidental, one may hope 
that the RGPEP formalism shall apply also to light hadrons as built 
from constituent quarks and massive gluons, the latter nearly 
decoupled after generating effective interactions for quarks on 
the way down in $\lambda$ toward 1/fm~\cite{GlazekWilson1998,GlazekMlynikAF}.
But even for heavy baryons alone, the effective oscillator picture 
provides simple wave functions that can be used in description of 
relativistic processes that involve heavy hadrons.

\begin{acknowledgements}
Mar\'ia G\'omez-Rocha acknowledges financial support from 
the European Centre for Theoretical Studies in Nuclear Physics 
and Related Areas, MINECO FPA2016-75654-C2-1-P and the 
European Commission under the Marie Sk\l odowska-Curie Action 
Cofund 2016 EU project 754446 -- Athenea3i. Stan G\l azek 
acknowledges support of ECT* during his visit in February 2017. 
\end{acknowledgements}

%%%%%
\appendix
%%%%%

%%%%%%%%%%%%%%%%%%%%%%%%%%
 \section{ Fits to masses of well-established quarkonia }
 \label{app:fits}
%%%%%%%%%%%%%%%%%%%%%%%%%%

\begin{figure}
\includegraphics[width=0.5\textwidth]{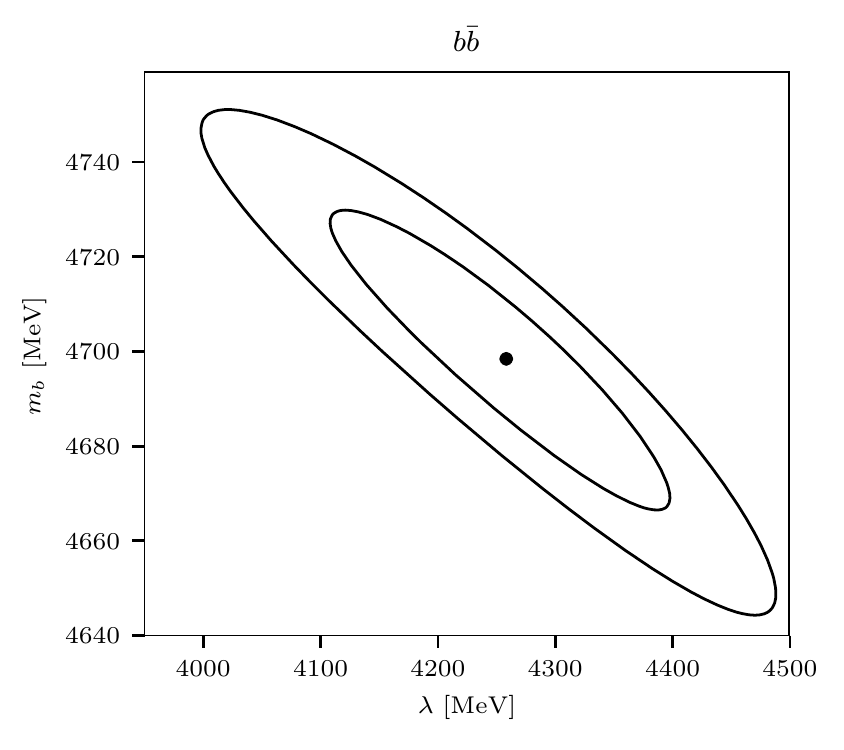}
\includegraphics[width=0.5\textwidth]{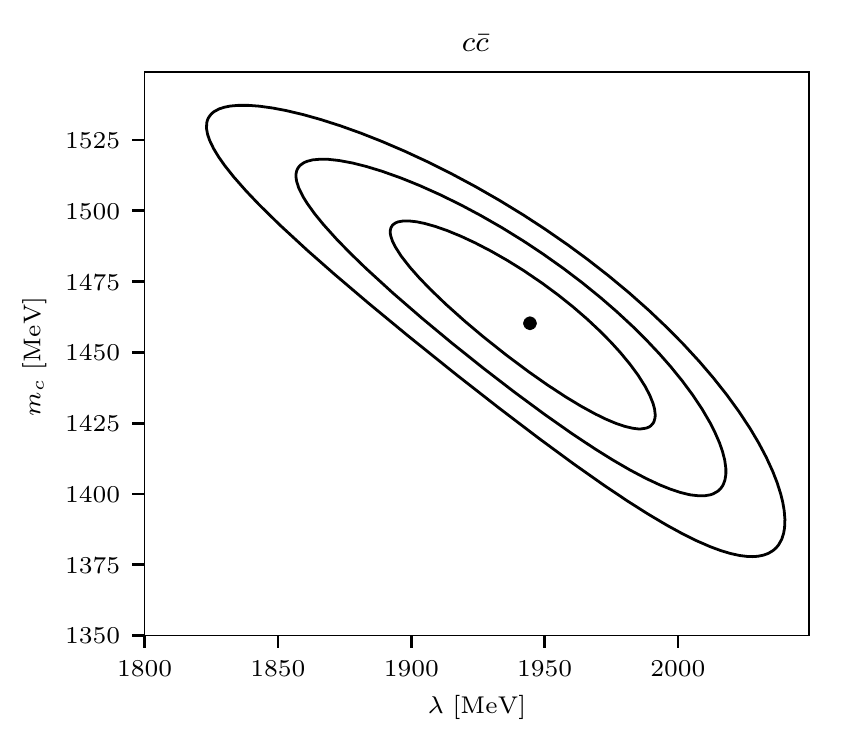}
\caption{Contour plots of the square root of mean squared 
deviation of the fit on $\lambda$-$m$ plane for bottomonium and 
charmonium. The plotted contours correspond to $30$ (for 
charmonium only), $40$ and $50$~MeV. Dots indicate best fits.}
\label{fig:fitmaps}
\end{figure}

Once the coupling constant $\alpha$ as a function
of $\lambda$ is set, the eigenvalues of Eq.~(\ref{EVEm})
estimated by evaluating expectation values of the Coulomb
terms in known eigenstates of the oscillator part of the 
effective Hamiltonian, 
\beq
E_{1S}
\es
\frac{3}{2}\omega - \frac{4}{3} \alpha \sqrt{\frac{2}{\pi\nu}}
\ ,
\label{e1S}
\\
E_{2S}
\es
\frac{7}{2}\omega - \frac{10}{9} \alpha \sqrt{\frac{2}{\pi\nu}}
\ ,
\label{e2S}
\\
E_{1P}
\es
\frac{5}{2}\omega - \frac{8}{9}\alpha \sqrt{\frac{2}{\pi\nu}}
\ ,
\eeq
with 
\beq
\omega
\es
\sqrt{\frac{\alpha(\lambda_{Q\bar Q})}{18\sqrt{2\pi}}} \frac{\lambda_{Q\bar Q}^3}{m^2}
\ ,
\quad \quad
\nu
\rs
\frac{1}{m \omega}
\ ,
\eeq
and $m = m_c$ or $m=m_b$, are used to evaluate 
corresponding masses from the formula
\beq
M
\es
2m \sqrt{1 + \frac{E}{m}}
\ .
\eeq
These are compared with data~\cite{PDG2016}.
Thus, $\Upsilon(1S)$, $\Upsilon(2S)$ and $\chi_{b1}(1P)$
are used to find best values $m_b$ and $\lambda_{b\bar b}$,
using $\chi^2$. We obtain,
\beq
m_b
\es
4698\text{ MeV}
\ ,
\quad \quad
\lambda_{b\bar b}
\rs
4258\text{ MeV}
\ .
\eeq
Using $\alpha(\lambda_{b\bar b}) = 0.2664$, we find that
$ \omega_{b\bar b} = 268.8$ MeV. Mean squared deviation (MSD) of 
the $b \bar b$ fit is $33$~MeV. Allowing for bigger MSD up to, 
say $50$~MeV, produces a set of acceptable values of $\lambda$ 
and $m_b$, illustrated in Fig.~\ref{fig:fitmaps}.
For charmonia, $J/\psi$, $\psi(2S)$ and $\chi_{c1}(1P)$
masses are used in the same way to find best values of
$m_c$ and $\lambda_{c\bar c}$, which turn out to be 
\beq
m_c
\es
1460\text{ MeV}
\ ,
\quad \quad
\lambda_{c\bar c}
\rs
1944\text{ MeV}
\ .
\eeq
Using $\alpha(\lambda_{c\bar c})=0.3926$, we obtain 
$\omega_{c\bar c} = 321.6$ MeV. The MSD of the $c \bar c$ 
fit is $22$~MeV, see Fig.~\ref{fig:fitmaps} for uncertainty 
of the fit. It is visible in Fig.~\ref{fig:fitmaps} that 
there exist functions $m_b(\lambda)$ and $m_c(\lambda)$ that 
one might introduce to obtain some window of stability of the 
fit accuracy, exceeding 10\% variation in $\lambda$. However, 
the pilot study appears too crude to us to believe that this 
stability already reflects the true behavior of quark masses 
in the theory, even though stability windows of that size 
naturally appear at order $\alpha$ in Hamiltonian matrix 
models with asymptotic freedom and bound states~\cite{GlazekWilson1998}. 

Two heavy mesons made of quarks $b$ and $c$ were observed,
$B_c$ and $B_c(2S)$~\cite{PDG2016}. Fits to quarkonia fix
masses of quarks, hence, the only free parameter is $\lambda_{bc}$,
which we fix by assuming Eq.~(\ref{lambdaQQ}). Without any
freedom left we plot in Fig.~\ref{fig:qqspectra} the spectrum
of $B_c$ using the same Eqs.~(\ref{e1S}) and (\ref{e2S}), but 
with
\beq
\omega^2
\es
\frac{\alpha(\lambda_{bc})}{18\sqrt{\pi}\, \mu_{bc}}
 \left(\frac{\lambda_{bc}^2}{\sqrt{m_b^2+m_c^2}}\right)^3
\ ,
\label{omegabc}
\\
\nu
\es
\frac{1}{2\mu_{bc}\omega}
\ ,
\\
\mu_{bc}
\es
\frac{m_b m_c}{m_b + m_c}
\ .
\eeq
The physical quarkonia masses are read from
\beq
M
\es
\left( m_b + m_c \right) \sqrt{1 + \frac{2E}{m_b + m_c}}
\ .
\eeq

%%%%%%%%%%%%%%%%%%%
 \section{ Parameters for heavy baryons }
 \label{app:predictions}
%%%%%%%%%%%%%%%%%%%

We choose parameter $\lambda$ for a baryon system by assuming
Eq.~(\ref{lambda3Q}) where $a$ and $b$
are given in Eqs.~(\ref{aQQbar}) and (\ref{bQQbar}). Values
of $\lambda$ are solutions to the following equations
\beq
\lambda_{bbb} \es  \sqrt{\alpha(\lambda_{bbb})}
\ \left( a\, m_b + b \right)
\ ,
\\
\lambda_{bbc} \es  \sqrt{\alpha(\lambda_{bbc})}
\ \left( a\, \frac{ 2 m_b + m_c}{3} + b \right)
\ ,
\\
\lambda_{ccb} \es  \sqrt{\alpha(\lambda_{ccb})}
\ \left( a\, \frac{ m_b + 2 m_c}{3} + b \right)
\ ,
\\
\lambda_{ccc} \es  \sqrt{\alpha(\lambda_{ccc})}
\ \left( a\, m_c + b \right)
\ .
\eeq
We obtain: 
\begin{align}
&\lambda_{bbb} \rs 4258\text{ MeV}\ ,
\quad
\alpha(\lambda_{bbb}) \rs 0.2664\ ,
\label{lambbb}
\\
&\lambda_{bbc} \rs 3514\text{ MeV}\ ,
\quad
\alpha(\lambda_{bbc}) \rs 0.2892\ ,
\\
&\lambda_{ccb} \rs 2746\text{ MeV}\ ,
\quad
\alpha(\lambda_{ccb}) \rs 0.3248\ ,
\\
&\lambda_{ccc} \rs 1944\text{ MeV}\ ,
\quad
\alpha(\lambda_{ccc}) \rs 0.3926\ .
\label{lamccc}
\end{align}
For readers' convenience, we also listed above 
the associated values of coupling constant.

%%%%%%%%%%%%%%%%%%%%%%%%%%%%%%%%%%%%%%%%%%%%%%%%%%%%%%%%%%%%%%%%
 \section{ Frequency diagram }
 \label{app:frequencies}
%%%%%%%%%%%%%%%%%%%%%%%%%%%%%%%%%%%%%%%%%%%%%%%%%%%%%%%%%%%%%%%%

\begin{figure}[h]
\centering
\includegraphics{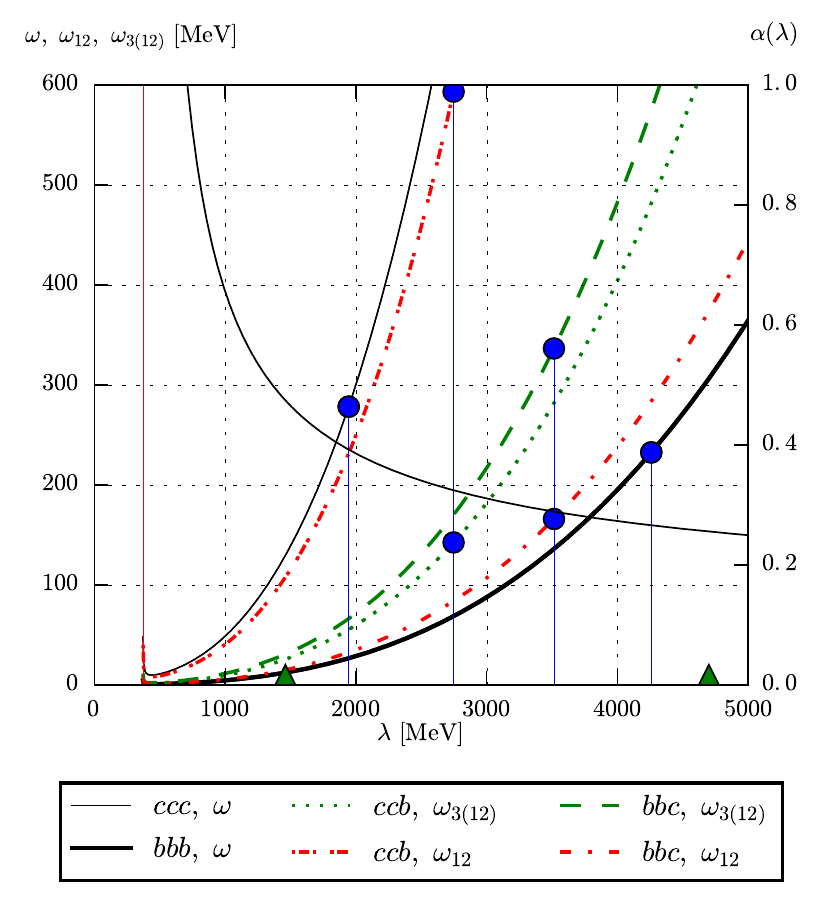}
\caption{ Dependence of the harmonic-oscillator frequencies
on the RGPEP scale $\lambda$ and quark masses (left axis),
and the dependence of $\alpha$ on $\lambda$ (right axis).
For detailed description of this figure content, see \ref{app:frequencies}. }
\label{fig:frequencies}
\end{figure}

Harmonic oscillator frequencies, Eqs.~(\ref{diquarkomega}) and
(\ref{omega3}), depend on quark masses and on the scale $\lambda$. 
Figure~\ref{fig:frequencies} shows the dependence of $\omega_{12}$ 
and $\omega_{3(12)}$ on $\lambda$ for four different choices of 
three quark masses that correspond to the systems $bbb$, $bbc$, 
$ccb$ and $ccc$. For $ccc$ and $bbb$ we have $\omega_{12} =
\omega_{3(12)} = \omega$. Blue vertical lines indicate the values of
$\lambda$s for baryons from Eqs.~(\ref{lambbb}) to (\ref{lamccc}).
They end on a higher of two blue dots. The dots show the values of
$\omega_{12}(\lambda)$ and $\omega_{3(12)}(\lambda)$ for a given 
system and for $\lambda$ adjusted to that system. The frequencies 
are,
\begin{align}
\omega_{bbb}
&=
232.8\text{ MeV,}
\\
\omega_{12,\, bbc}
&=
166.2\text{ MeV,}
\quad
\omega_{3(12),\, bbc}
=
336.7\text{ MeV,}
\\
\omega_{12,\, ccb}
&=
593.5\text{ MeV,}
\quad
\omega_{3(12),\, ccb}
=
142.7\text{ MeV,}
\\
\omega_{ccc}
&=
278.5\text{ MeV.}
\end{align}
On Fig.~\ref{fig:frequencies}, there are also two green triangles 
at the bottom of the plot. They indicate the values of $m_b$ and $m_c$.
Furthermore, Figure~\ref{fig:frequencies} presents also $\alpha(\lambda)$
given in Eq.~(\ref{alphaQCD}) with a decreasing black line.
Red vertical line shows the asymptote of $\alpha(\lambda)$
curve at $\lambda=\Lambda_{QCD}$.

%%%%%%%%%%%%%%%%%%
 \section{ Wave functions for baryons }
 \label{app:wavefunctions}
%%%%%%%%%%%%%%%%%%

The unperturbed harmonic oscillator basis for baryons 
is constructed from products of wave functions of two 
harmonic oscillators associated with relative motion 
of particles $1$ and $2$ (with momentum $\vec K_{12}$), 
and relative motion of particle $3$ of momentum $\vec Q_3$ 
with respect to the pair $12$.

Eigenfunctions for relative motion of two particles with
reduced mass $\mu$ interacting with harmonic oscillator
force characterized by frequency $\omega$ are
\begin{align}
\psi_{klm}(p,\theta,\phi)
=
N_{kl}\  e^{-\nu p^2}\  L_k^{(l+1/2)}(2\nu p^2)
\ p^l Y_{lm}(\theta,\phi)
\ ,
\end{align}
where $\nu=1/(2\mu\omega)$, $k$ is the radial excitation number,
$l$ is the orbital angular momentum number of the state, $m$ is
the projection of angular momentum on the $z$ axis, $Y_{lm}$ are
spherical harmonics~\cite{PDG2016} and $L_k^a(x)$ are generalized
Laguerre polynomials. The first two polynomials are $L_0^a(x)=1$,
$L_1^a(x)=1+a-x$. The normalization factors are
\beq
N_{kl}
\es
\sqrt{\sqrt{\frac{2\nu^3}{\pi}}
\frac{2^{k+2l+3} k! \nu^l}{(2k+2l+1)!!}} (2\pi)^{3/2}
\ ,
\eeq
so that
$
\int\frac{d^3p}{(2\pi)^3} \psi_{klm}^* \psi_{k'l'm'}
=
\delta_{kk'}\delta_{ll'}\delta_{mm'}
\ .
$
Finally, the energies are
\beq
E
\es \omega\left(2k+l+\frac{3}{2}\right)
\ .
\eeq
We use a convenient notation for products of wave 
functions of two harmonic oscillators,
\begin{align}
\ket{(k_{12}+1)(l_{12})_{m_{12}} (k_{3(12)}+1)(l_{3(12)})_{m_{3(12)}} }
=
\ket{\psi_{k_{12} l_{12} m_{12}}}
\ket{\psi_{k_{3(12)} l_{3(12)} m_{3(12)}}}
\ ,
\end{align}
where index $12$ corresponds to harmonic oscillator between
$1$ and $2$, with $\nu_{12} = 1/(2\mu_{12}\omega_{12})$,
and index $3(12)$ corresponds to harmonic oscillator between
$3$ and $12$, with $\nu_{3(12)} = 1/(2\mu_{3(12)}\omega_{3(12)})$.
For example, the ground state is $\ket{1S_01S_0}\equiv\ket{1S1S}$,
while $\ket{1P_12S_0}\equiv\ket{1P_12S}$ is the state with
harmonic oscillator between $1$ and $2$ excited to the first
orbital excitation with angular momentum projection on $z$-axis
equal $1$ and the harmonic oscillator between $3$ and $12$
being radially excited. The quantum numbers $m_{12}$ and
$m_{3(12)}$ are omitted below, unless they are relevant.

We consider states whose excitation energies are at most $2\omega_{12}$
or $2\omega_{3(12)}$ or $\omega_{12}+\omega_{3(12)}$. That is, 
we consider states $1S1S$, $1P1S$, $1S1P$, $2S1S$, $1S2S$, $1D1S$, $1S1D$
and $1P1P$. The total angular momentum of a baryon is conserved,
therefore, each state of the basis should have definite orbital
angular momentum $L$. Since states $\ket{1P_{m_{12}}1P_{m_{3(12)}}}$
do not have definite angular momentum, we introduce instead 
the following states with angular momenta $L=2,1$ and $0$, 
respectively,
\begin{align}
\ket{2,+2}
&=
\ket{1P_1 1P_1}
\ ,
\label{wf1P1PL2}
\quad
\ket{2,+1} = \dots
\\
\ket{1,+1}
&=
\frac{1}{\sqrt{2}} \ket{1P_1 1P_0} - \frac{1}{\sqrt{2}} \ket{1P_0 1P_1}
\ ,
\quad
\ket{1,0} = \dots
\\
\ket{0,\phantom{+}0}
&=
\frac{1}{\sqrt{3}}
\left( \phantom{\frac{}{}}
  \ket{1P_1 1P_{-1}}
- \ket{1P_0 1P_0}
+ \ket{1P_{-1} 1P_1}
\phantom{\frac{}{}}\right)
\ ,
\label{wf1P1PL0}
\end{align}
where only the highest $L_z$ state is written explicitly.
The construction of these states is done in accordance
with the rules of adding angular momenta in quantum 
mechanics and we use convention defined in~\cite{PDG2016}
in the tables of Clebsch-Gordan coefficients.

Because quarks have spin, we also need to construct the 
spin wave functions. We define spin-$3/2$ quadruplet, 
which is fully symmetric with respect to exchange of 
any pair of quarks,
\beq
\ket{+\frac{3}{2}}
\es
\ket{\up\up\up}
\ ,
\quad
\ket{+\frac{1}{2}} \rs \dots
\ ;
\eeq
spin-$1/2$ doublet, which we call $(1/2)_S$ and which is $12$-symmetric,
\beq
\ket{+\frac{1}{2}S}
\es
\sqrt{\frac{2}{3}}
\left(
\phantom{+} \ket{\up\up\down} - \frac{1}{2} \ket{\up\down\up} - \frac{1}{2} \ket{\down\up\up}
\right)
\ ,
\\
\ket{-\frac{1}{2}S}
\es
\sqrt{\frac{2}{3}}
\left(
- \ket{\down\down\up} + \frac{1}{2} \ket{\down\up\down} + \frac{1}{2} \ket{\up\down\down}
\right)
\ ;
\eeq
and another spin-$1/2$ doublet, which we call $(1/2)_A$ and which is
$12$-antisymmetric,
\beq
\ket{+\frac{1}{2}A}
\es
\sqrt{\frac{1}{2}}
\left(
\ket{\up\down\up} - \ket{\down\up\up}
\right)
\ ,
\\
\ket{-\frac{1}{2}A}
\es
\sqrt{\frac{1}{2}}
\left(
\ket{\up\down\down} - \ket{\down\up\down}
\right)
\ .
\eeq

For completeness, we describe our construction of states 
of definite total angular momentum. First consider $ccb$ 
and $bbc$ systems, where quarks $1$ and $2$ are identical 
and $3$ is different. Because quarks $1$ and $2$ are identical, 
the total spin-momentum wave function has to be $12$-symmetric; 
color-singlet wave function is antisymmetric. Therefore, one 
must add orbital angular momentum and spin respecting 
the Pauli exclusion principle for fermions. For example, the
states $\ket{1P1S}$ are $12$-antisymmetric and to obtain 
$12$-symmetric spin-momentum wave function we can 
combine them only with spin $(1/2)_A$, which is also 
$12$-antisymmetric. States $\ket{1S1P}$ are $12$-symmetric 
and we can combine them only with spin $3/2$ and $(1/2)_S$.
The list of possible states is summarized  in Table~\ref{tab:states}.
\begin{table}
\centering
\caption{ Summary of oscillator basis states for systems $ccb$ and $bbc$.}
\label{tab:states}
\begin{tabular*}{\columnwidth}{@{\extracolsep{\fill}}lll@{}}
\hline
States & $L \otimes S$ & J \\
\hline
 1S1S & $0 \otimes \frac{3}{2}$ & ${\frac{3}{2}}^+$ \\
      & $0 \otimes \left(\frac{1}{2}\right)_S$ & ${\frac{1}{2}}^+$ \\
\hline
 1P1S & $1 \otimes \left(\frac{1}{2}\right)_A$ & ${\frac{3}{2}}^- \oplus {\frac{1}{2}}^-$ \\
\hline
 1S1P & $1 \otimes \frac{3}{2}$ & ${\frac{5}{2}}^- \oplus {\frac{3}{2}}^- \oplus {\frac{1}{2}}^-$ \\
      & $1 \otimes \left(\frac{1}{2}\right)_S$ & ${\frac{3}{2}}^- \oplus {\frac{1}{2}}^-$ \\
\hline
 1D1S & $2 \otimes \frac{3}{2}$ & ${\frac{7}{2}}^+ \oplus {\frac{5}{2}}^+ \oplus {\frac{3}{2}}^+ \oplus {\frac{1}{2}}^+$ \\
      & $2 \otimes \left(\frac{1}{2}\right)_S$ & ${\frac{5}{2}}^+ \oplus {\frac{3}{2}}^+$ \\
\hline
 1S1D & $2 \otimes \frac{3}{2}$ & ${\frac{7}{2}}^+ \oplus {\frac{5}{2}}^+ \oplus {\frac{3}{2}}^+ \oplus {\frac{1}{2}}^+$ \\
      & $2 \otimes \left(\frac{1}{2}\right)_S$ & ${\frac{5}{2}}^+ \oplus {\frac{3}{2}}^+$ \\
\hline
 1P1P & $2 \otimes \left(\frac{1}{2}\right)_A$ & ${\frac{5}{2}}^+ \oplus {\frac{3}{2}}^+$ \\
      & $1 \otimes \left(\frac{1}{2}\right)_A$ & ${\frac{3}{2}}^+ \oplus {\frac{1}{2}}^+$ \\
      & $0 \otimes \left(\frac{1}{2}\right)_A$ & ${\frac{1}{2}}^+$ \\
\hline
\end{tabular*}
\end{table}
To obtain the explicit formulas for the wave functions, 
we use Clebsch-Gordan tables, as in Eqs.~(\ref{wf1P1PL2}) to (\ref{wf1P1PL0}).

%%%%%%%%%%%%%%%%%%%%%%%%%%%%%%%%%%%%%%%%%%%%%%%%
 \subsection{ Symmetric wave functions }
%%%%%%%%%%%%%%%%%%%%%%%%%%%%%%%%%%%%%%%%%%%%%%%%

In the case of three identical quarks, we need to use fully
symmetric wave functions. One can symmetrize the wave
functions given above. The ground state $1S1S$ wave
function is fully symmetric in momentum, and we can 
combine it only with spin $3/2$. By the way, symmetrization 
of $1S1S$ with $(1/2)_S$ gives zero. In this case the wave 
function is the same as in the case of only two quarks being 
identical,
\beq
\ket{0\omega,\frac{3}{2}^+,J_z}
\es
\ket{1S1S}\ket{J_z}
\ ,
\eeq
where $J_z=+3/2,+1/2,-1/2,-3/2$ is the projection of baryon 
spin on $z$-axis.

After symmetrization of the oscillator once-excited states $1P1S$ 
and $1S1P$, one is left with only two linearly independent multiplets 
of states, whose wave functions are (we write only the highest $J_z$ 
state in each multiplet, more information is available in
Table~\ref{tab:statessymmetrized})
\beq
\ket{1\omega,\frac{3}{2}^-,+\frac{3}{2}}
\es
  \frac{1}{\sqrt{2}}
  \ket{1P_1 1S} \ket{+\frac{1}{2}A}
- \frac{1}{\sqrt{2}}
  \ket{1S 1P_1} \ket{+\frac{1}{2}S}
\ ,
% \end{align}
\\
% \begin{align}
\ket{1\omega,\frac{1}{2}^-,+\frac{1}{2}}
\es
- \frac{1}{\sqrt{6}}
  \ket{1P_0 1S} \ket{+\frac{1}{2}A}
+ \frac{1}{\sqrt{3}}
  \ket{1P_1 1S} \ket{-\frac{1}{2}A}
\np
  \frac{1}{\sqrt{6}}
  \ket{1S 1P_0} \ket{+\frac{1}{2}S}
- \frac{1}{\sqrt{3}}
  \ket{1S 1P_1} \ket{-\frac{1}{2}S}
\ .
\eeq

The symmetrization of a band of twice-excited oscillator states, $2S1S$,
$1S2S$, $1D1S$, $1S1D$ and $1P1P$ reduces the number of linearly
independent multiplets from 21 to 8 (compare Tables~\ref{tab:states}
and \ref{tab:statessymmetrized}). Highest $J_z$ states in each
multiplet are
\beq
\ket{A_{\frac{3}{2}^+},+\frac{3}{2}}
\es
\ket{2S 1S}_+ \ket{+\frac{3}{2}}
\ ,
\\
\ket{B_{\frac{1}{2}^+},+\frac{1}{2}}
\es
  \frac{1}{\sqrt{2}}
  \ket{2S 1S}_- \ket{+\frac{1}{2}S}
- \frac{1}{\sqrt{2}}
  \ket{0,0} \ket{+\frac{1}{2}A}
\ ,
\\
\ket{C_{\frac{7}{2}^+},+\frac{7}{2}}
\es
\ket{1D_2 1S}_+ \ket{+\frac{3}{2}}
\ ,
\\
\ket{C_{\frac{5}{2}^+},+\frac{5}{2}}
\es
- \sqrt{\frac{3}{7}}
  \ket{1D_1 1S}_+ \ket{+\frac{3}{2}}
+ \sqrt{\frac{4}{7}}
  \ket{1D_2 1S}_+ \ket{+\frac{1}{2}}
\ ,
\\
\ket{C_{\frac{3}{2}^+},+\frac{3}{2}}
\es
  \sqrt{\frac{1}{5}}
  \ket{1D_0 1S}_+ \ket{+\frac{3}{2}}
- \sqrt{\frac{2}{5}}
  \ket{1D_1 1S}_+ \ket{+\frac{1}{2}}
+
  \sqrt{\frac{2}{5}}
  \ket{1D_2 1S}_+ \ket{-\frac{1}{2}}
\ ,
\nn
\eeq
\beq
\ket{C_{\frac{1}{2}^+},+\frac{1}{2}}
\es
- \sqrt{\frac{1}{10}}
  \ket{1D_{-1} 1S}_+ \ket{+\frac{3}{2}}
+ \sqrt{\frac{1}{5}}
  \ket{1D_0 1S}_+ \ket{+\frac{1}{2}}
\nm
  \sqrt{\frac{3}{10}}
  \ket{1D_1 1S}_+ \ket{-\frac{1}{2}}
+ \sqrt{\frac{2}{5}}
  \ket{1D_2 1S}_+ \ket{-\frac{3}{2}}
\ ,
\\
\ket{D_{\frac{5}{2}^+},+\frac{5}{2}}
\es
  \frac{1}{\sqrt{2}}
  \ket{1D_2 1S}_- \ket{+\frac{1}{2}S}
- \frac{1}{\sqrt{2}}
  \ket{2,+2} \ket{+\frac{1}{2}A}
\ ,
\label{symmetrizedD52}
\\
\ket{D_{\frac{3}{2}^+},+\frac{3}{2}}
\es
- \sqrt{\frac{1}{10}}
  \ket{1D_1 1S}_- \ket{+\frac{1}{2}S}
+ \sqrt{\frac{2}{5}}
  \ket{1D_2 1S}_- \ket{-\frac{1}{2}S}
\np \sqrt{\frac{1}{10}}
  \ket{2,+1} \ket{+\frac{1}{2}A}
- \sqrt{\frac{2}{5}}
  \ket{2,+2} \ket{-\frac{1}{2}A}
\ ,
\label{symmetrizedD32}
\eeq
where
\beq
\ket{2S1S}_\pm
\es
\frac{\ket{2S1S} \pm \ket{1S2S}}{\sqrt{2}}
\ ,
\\
\ket{1D_m1S}_\pm
\es
\frac{\ket{1D_m1S} \pm \ket{1S1D_m}}{\sqrt{2}}
\ .
\eeq
States with $J_z$ different than the highest $J_z$
available in the multiplet can be constructed according
to Table~\ref{tab:statessymmetrized}.
\begin{table}[h]
\begin{center}
\caption{ Summary of states for systems $ccc$ and $bbb$. 
For example, $1D1S_- \otimes \left(\frac{1}{2}\right)_S$ 
means that we use $\ket{1D_m 1S}_-$ states and $(1/2)_S$ 
spin states to obtain one of $J=5/2$ or $J=3/2$ states 
according to the rules of adding angular momenta, {\it i.e.}, 
using the Clebsch-Gordan coefficients. $1P1P_{L=2} \otimes 
\left(\frac{1}{2}\right)_A$ means that we take $L=2$
states, given in Eq.~(\ref{wf1P1PL2}), and $(1/2)_A$ spin 
states to obtain a state with the same quantum numbers. 
We then subtract the latter from the former, as indicated in the table, 
and normalize the result to obtain the final expression, 
such as in Eqs.~(\ref{symmetrizedD52}) or (\ref{symmetrizedD32}), 
where the states with $J=5/2$, $J_z=+5/2$ and $J=3/2$, $J_z=+3/2$
are written explicitly. Our prescription differs in sign from the 
prescriptions known in the literature~\cite{Isgur:1978wd,Capstick:2000qj},
because our momentum $\vec Q_3$ is a momentum of quark $3$ 
with respect to pair $12$, instead of pair $12$ with respect to quark
$3$. }
\label{tab:statessymmetrized}
\begin{tabular*}{\columnwidth}{@{\extracolsep{\fill}}lll@{}}
\hline
 States & Wave functions & Baryons \\
\hline
 $0\omega$ & $1S1S \otimes \frac{3}{2}$ & ${\frac{3}{2}}^+$ \\
\hline
 $1\omega$ & $1P1S \otimes \left(\frac{1}{2}\right)_A - 1S1P \otimes \left(\frac{1}{2}\right)_S$ & ${\frac{3}{2}}^- \oplus {\frac{1}{2}}^-$ \\
\hline
 A    & $2S1S_+ \otimes \frac{3}{2}$ & ${\frac{3}{2}}^+$ \\
% \hline
 B    & $2S1S_- \otimes \left(\frac{1}{2}\right)_S - 1P1P_{L=0} \otimes \left(\frac{1}{2}\right)_A$ & ${\frac{1}{2}}^+$ \\
% \hline
 C    & $1D1S_+ \otimes \frac{3}{2}$ & ${\frac{7}{2}}^+ \oplus \dots \oplus {\frac{1}{2}}^+$ \\
% \hline
 D    & $1D1S_- \otimes \left(\frac{1}{2}\right)_S - 1P1P_{L=2} \otimes \left(\frac{1}{2}\right)_A$ & ${\frac{5}{2}}^+ \oplus {\frac{3}{2}}^+$ \\
\hline
\end{tabular*}
\end{center}
\end{table}

%%%%%%%%%%%%%
 \section{ Baryon masses }
 \label{app:masses}
%%%%%%%%%%%%%

%%%%%%%%%%%%%%%
\subsection{ States $ccb$ and $bbc$ }
%%%%%%%%%%%%%%%

Baryon masses are given by Eq.~(\ref{massbaryon}), where
\beq
E
\es
  \omega_{12}    \left(2k_{12} + l_{12} + \frac{3}{2} \right)
+ \omega_{3(12)} \left(2k_{3(12)} + l_{3(12)} + \frac{3}{2} \right)
+ V
\ ,
\label{energybaryon}
\eeq
and $V = \bra{\cdot} \, \hat V_C \, \ket{\cdot}$ is the expectation
value of Coulomb interaction in the harmonic oscillator eigenstates.
For example,
\beq
E_{1P1P}^{L=0}
\es
  \frac{5}{2} \omega_{12}
+ \frac{5}{2} \omega_{3(12)} + V_{1P1P}^{L=0}
\ ,
\eeq
where $V_{1P1P}^{L=0} = \bra{0,0} \, \hat V_C \, \ket{0,0}$.
We define,
\beq
V
\es
- \frac{2}{3} \alpha \sqrt{\frac{2}{\pi\nu_{12}}} \tilde V
\label{Vbaryon}
\eeq
and
\beq
x \es \frac{4\nu_{3(12)}}{\nu_{12}}
\ .
\label{xbaryon}
\eeq
We list the Coulomb interaction expectation values for 
$ccb$ and $bbc$ states in Fig.~\ref{fig:QQQ}. 
\beq
\tilde V_{1S1S}
\es
1 + \frac{4}{\sqrt{1+x}}
\ ,
% \eeq
\\
% \beq
\tilde V_{1P1S}
\es
\frac{2}{3} + \frac{4 \left(3 x+2\right)}{3 (1+x)^{3/2}}
\ ,
% \eeq
\\
% \beq
\tilde V_{1S1P}
\es
1 + \frac{4 (2 x+3)}{3 (1+x)^{3/2}}
\ ,
% \eeq
\\
% \beq
\tilde V_{1P1P}^{L=0}
\es
\frac{2}{3} + \frac{4(2x^2+7x+2)}{3(1+x)^{5/2}}
\ ,
% \eeq
\\
% \beq
\tilde V_{1P1P}^{L=1}
\es
\frac{2}{3} + \frac{8}{3\sqrt{x+1}}
\ ,
% \eeq
\\
% \beq
\tilde V_{1P1P}^{L=2}
\es
\frac{2}{3} + \frac{8 \left(5 x^2+13 x+5\right)}{15(x+1)^{5/2}}
\ ,
% \eeq
\\
% \beq
\tilde V_{1D1S}
\es
\frac{8}{15} + \frac{4 \left(15 x^2+ 20 x + 8\right)}{15(1+x)^{5/2}}
\ ,
% \eeq
\\
% \beq
\tilde V_{1S1D}
\es
1 + \frac{4 \left(8 x^2+20 x+15\right)}{15 (1+x)^{5/2}}
\ ,
% \eeq
\\
% \beq
\tilde V_{2S1S}
\es
\frac{5}{6} + \frac{2 \left(6 x^2+8 x+5\right)}{3 (1+x)^{5/2}}
\ ,
% \eeq
\\
% \beq
\tilde V_{1S2S}
\es
1 + \frac{2 \left(5 x^2+8 x+6\right)}{3 (1+x)^{5/2}}
\ .
\eeq

%%%%%%%%%%%%%%%%%%
\subsection{ States $ccc$ and $bbb$ }
%%%%%%%%%%%%%%%%%%

For baryons $ccc$ and $bbb$, we also make use of 
Eqs.~(\ref{energybaryon}), (\ref{Vbaryon}) and (\ref{xbaryon}). 
Formulas for ground states and once orbitally excited states do 
not change. For identical quarks $x=3$, and
\beq
\tilde V_{1S1S}
\es
3
\ ,
\\
\tilde V_{1\omega}
\es
\frac{5}{2}
\ .
\eeq
Energies for states $A$, $B$, $C$ and $D$ need to 
be evaluated separately. We have
\beq
\tilde V_{A}
\es
\frac{11}{4}
\ ,
\\
\tilde V_{B}
\es
\frac{19}{8}
\ ,
\\
\tilde V_{C}
\es
\frac{23}{10}
\ ,
\\
\tilde V_{D}
\es
\frac{43}{20}
\ .
\eeq

% BibTeX users please use one of
%\bibliographystyle{spbasic}      % basic style, author-year citations
%\bibliographystyle{spmpsci}      % mathematics and physical sciences
%\bibliographystyle{spphys}       % APS-like style for physics
%\bibliography{RGPEPrefs.bib}   % name your BibTeX data base

\begin{thebibliography}{10}
\providecommand{\url}[1]{{#1}}
\providecommand{\urlprefix}{URL }
\expandafter\ifx\csname urlstyle\endcsname\relax
  \providecommand{\doi}[1]{DOI \discretionary{}{}{}#1}\else
  \providecommand{\doi}{DOI \discretionary{}{}{}\begingroup
  \urlstyle{rm}\Url}\fi

\bibitem{Capstick:2000qj}
S.~Capstick, W.~Roberts, Prog. Part. Nucl. Phys. \textbf{45}, S241 (2000).
\newblock \doi{10.1016/S0146-6410(00)00109-5}

\bibitem{Capstick:1986bm}
S.~Capstick, N.~Isgur, Phys. Rev. \textbf{D34}, 2809 (1986).
\newblock \doi{10.1103/PhysRevD.34.2809}

\bibitem{QQbarRGPEP}
S.D. Głazek, M.~Gómez-Rocha, J.~More, K.~Serafin, Phys. Lett.
  \textbf{B773}(9), 172 (2017).
\newblock \doi{10.1016/j.physletb.2017.08.018}

\bibitem{tHooft:1973mfk}
G.~'t~Hooft, Nucl. Phys. \textbf{B61}, 455 (1973).
\newblock \doi{10.1016/0550-3213(73)90376-3}

\bibitem{Wilsonetal}
K.G. Wilson, T.S. Walhout, A.~Harindranath, W.M. Zhang, R.J. Perry, S.D.
  Glazek, Phys. Rev. \textbf{D49}, 6720 (1994).
\newblock \doi{10.1103/PhysRevD.49.6720}

\bibitem{Parisi:1980jy}
G.~Parisi, R.~Petronzio, Phys. Lett. \textbf{94B}, 51 (1980).
\newblock \doi{10.1016/0370-2693(80)90822-9}

\bibitem{Cornwall:1981zr}
J.M. Cornwall, Phys. Rev. \textbf{D26}, 1453 (1982).
\newblock \doi{10.1103/PhysRevD.26.1453}

\bibitem{Aguilar:2017dco}
A.C. Aguilar, D.~Binosi, C.T. Figueiredo, J.~Papavassiliou, Eur. Phys. J.
  \textbf{C78}(3), 181 (2018).
\newblock \doi{10.1140/epjc/s10052-018-5679-2}

\bibitem{PDG2016}
C.~Patrignani, et~al., Chin. Phys. \textbf{C40}(10), 100001 (2016).
\newblock \doi{10.1088/1674-1137/40/10/100001}

\bibitem{Eichten:1978tg}
E.~Eichten, K.~Gottfried, T.~Kinoshita, K.D. Lane, T.M. Yan, Phys. Rev.
  \textbf{D17}, 3090 (1978).
\newblock \doi{10.1103/PhysRevD.17.3090}.
\newblock [Erratum: Phys. Rev.D21,313(1980)]

\bibitem{pRGPEP}
S.D. Glazek, Acta Phys. Polon. \textbf{B43}, 1843 (2012).
\newblock \doi{10.5506/APhysPolB.43.1843}

\bibitem{Semay:1997ys}
C.~Semay, B.~Silvestre-Brac, Nucl. Phys. \textbf{A618}, 455 (1997).
\newblock \doi{10.1016/S0375-9474(97)00060-2}

\bibitem{Juge:2002br}
K.J. Juge, J.~Kuti, C.~Morningstar, Phys. Rev. Lett. \textbf{90}, 161601
  (2003).
\newblock \doi{10.1103/PhysRevLett.90.161601}

\bibitem{Takahashi:2002it}
T.T. Takahashi, H.~Suganuma, Phys. Rev. Lett. \textbf{90}, 182001 (2003).
\newblock \doi{10.1103/PhysRevLett.90.182001}

\bibitem{Dirac1949}
P.A.M. Dirac, Rev. Mod. Phys. \textbf{21}, 392 (1949).
\newblock \doi{10.1103/RevModPhys.21.392}

\bibitem{Lepage:1980fj}
G.P. Lepage, S.J. Brodsky, Phys. Rev. \textbf{D22}, 2157 (1980).
\newblock \doi{10.1103/PhysRevD.22.2157}

\bibitem{Brodsky-Pauli-Pinsky}
S.J. Brodsky, H.C. Pauli, S.S. Pinsky, Phys. Rept. \textbf{301}, 299 (1998).
\newblock \doi{10.1016/S0370-1573(97)00089-6}

\bibitem{SRG1}
S.D. Glazek, K.G. Wilson, Phys. Rev. \textbf{D48}, 5863 (1993).
\newblock \doi{10.1103/PhysRevD.48.5863}

\bibitem{Casher:1974xd}
A.~Casher, L.~Susskind, Phys. Rev. \textbf{D9}, 436 (1974).
\newblock \doi{10.1103/PhysRevD.9.436}

\bibitem{Maris:1997hd}
P.~Maris, C.D. Roberts, P.C. Tandy, Phys. Lett. \textbf{B420}, 267 (1998).
\newblock \doi{10.1016/S0370-2693(97)01535-9}

\bibitem{Brodsky:2010xf}
S.J. Brodsky, C.D. Roberts, R.~Shrock, P.C. Tandy, Phys. Rev. \textbf{C82},
  022201 (2010).
\newblock \doi{10.1103/PhysRevC.82.022201}

\bibitem{GlazekCondensatesAPP}
S.D. Glazek, Acta Phys. Polon. \textbf{B42}, 1933 (2011).
\newblock \doi{10.5506/APhysPolB.42.1933}

\bibitem{Brodsky:2012ku}
S.J. Brodsky, C.D. Roberts, R.~Shrock, P.C. Tandy, Phys. Rev. \textbf{C85},
  065202 (2012).
\newblock \doi{10.1103/PhysRevC.85.065202}

\bibitem{AF}
M.~Gómez-Rocha, S.D. Głazek, Phys. Rev. \textbf{D92}(6), 065005 (2015).
\newblock \doi{10.1103/PhysRevD.92.065005}

\bibitem{Wilson1970}
K.G. Wilson, Phys. Rev. \textbf{D2}, 1438 (1970).
\newblock \doi{10.1103/PhysRevD.2.1438}

\bibitem{TrawinskiConfinement}
A.P. Trawiński, S.D. Głazek, S.J. Brodsky, G.F. de~Téramond, H.G. Dosch,
  Phys. Rev. \textbf{D90}(7), 074017 (2014).
\newblock \doi{10.1103/PhysRevD.90.074017}

\bibitem{GlazekSchadenCondensate}
S.D. Głazek, M.~Schaden, Phys. Lett. \textbf{B198}, 42 (1987).
\newblock \doi{10.1016/0370-2693(87)90154-7}

\bibitem{GlazekWilson1998}
S.D. Glazek, K.G. Wilson, Phys. Rev. \textbf{D57}, 3558 (1998).
\newblock \doi{10.1103/PhysRevD.57.3558}

\bibitem{GlazekMlynikAF}
S.D. Glazek, J.~Mlynik, Phys. Rev. \textbf{D67}, 045001 (2003).
\newblock \doi{10.1103/PhysRevD.67.045001}

\bibitem{Gomez-Rocha:2016cji}
M.~Gómez-Rocha, T.~Hilger, A.~Krassnigg, Phys. Rev. \textbf{D93}(7), 074010
  (2016).
\newblock \doi{10.1103/PhysRevD.93.074010}

\bibitem{Bjorken:1985ei}
J.D. Bjorken, AIP Conf. Proc. \textbf{132}, 390 (1985).
\newblock \doi{10.1063/1.35379}

\bibitem{Tsuge:1985ei}
M.~Tsuge, T.~Morii, J.~Morishita, Mod. Phys. Lett. \textbf{A1}, 131 (1986).
\newblock \doi{10.1142/S0217732386000191}.
\newblock [Erratum: Mod. Phys. Lett.A2,283(1987)]

\bibitem{SilvestreBrac:1996bg}
B.~Silvestre-Brac, Few Body Syst. \textbf{20}, 1 (1996).
\newblock \doi{10.1007/s006010050028}

\bibitem{Jia:2006gw}
Y.~Jia, JHEP \textbf{10}, 073 (2006).
\newblock \doi{10.1088/1126-6708/2006/10/073}

\bibitem{Martynenko:2007je}
A.P. Martynenko, Phys. Lett. \textbf{B663}, 317 (2008).
\newblock \doi{10.1016/j.physletb.2008.04.030}

\bibitem{Roberts:2007ni}
W.~Roberts, M.~Pervin, Int. J. Mod. Phys. \textbf{A23}, 2817 (2008).
\newblock \doi{10.1142/S0217751X08041219}

\bibitem{Vijande:2015faa}
J.~Vijande, A.~Valcarce, H.~Garcilazo, Phys. Rev. \textbf{D91}(5), 054011
  (2015).
\newblock \doi{10.1103/PhysRevD.91.054011}

\bibitem{Giannuzzi:2009gh}
F.~Giannuzzi, Phys. Rev. \textbf{D79}, 094002 (2009).
\newblock \doi{10.1103/PhysRevD.79.094002}

\bibitem{Ghalenovi:2014swa}
Z.~Ghalenovi, A.A. Rajabi, S.x. Qin, D.H. Rischke, Mod. Phys. Lett.
  \textbf{A29}, 1450106 (2014).
\newblock \doi{10.1142/S0217732314501065}

\bibitem{Shah:2017jkr}
Z.~Shah, A.K. Rai, Eur. Phys. J. \textbf{A53}(10), 195 (2017).
\newblock \doi{10.1140/epja/i2017-12386-2}

\bibitem{Ponce:1978gk}
W.~Ponce, Phys. Rev. \textbf{D19}, 2197 (1979).
\newblock \doi{10.1103/PhysRevD.19.2197}

\bibitem{Hasenfratz:1980ka}
P.~Hasenfratz, R.R. Horgan, J.~Kuti, J.M. Richard, Phys. Lett. \textbf{94B},
  401 (1980).
\newblock \doi{10.1016/0370-2693(80)90906-5}

\bibitem{Bernotas:2008bu}
A.~Bernotas, V.~Simonis, Lith. J. Phys. \textbf{49}, 19 (2009).
\newblock \doi{10.3952/lithjphys.49110}

\bibitem{Wei:2015gsa}
K.W. Wei, B.~Chen, X.H. Guo, Phys. Rev. \textbf{D92}(7), 076008 (2015).
\newblock \doi{10.1103/PhysRevD.92.076008}

\bibitem{Wei:2016jyk}
K.W. Wei, B.~Chen, N.~Liu, Q.Q. Wang, X.H. Guo, Phys. Rev. \textbf{D95}(11),
  116005 (2017).
\newblock \doi{10.1103/PhysRevD.95.116005}

\bibitem{Zhang:2009re}
J.R. Zhang, M.Q. Huang, Phys. Lett. \textbf{B674}, 28 (2009).
\newblock \doi{10.1016/j.physletb.2009.02.056}

\bibitem{Wang:2011ae}
Z.G. Wang, Commun. Theor. Phys. \textbf{58}, 723 (2012).
\newblock \doi{10.1088/0253-6102/58/5/17}

\bibitem{Aliev:2012tt}
T.M. Aliev, K.~Azizi, M.~Savci, JHEP \textbf{04}, 042 (2013).
\newblock \doi{10.1007/JHEP04(2013)042}

\bibitem{Aliev:2014lxa}
T.M. Aliev, K.~Azizi, M.~Savcı, J. Phys. \textbf{G41}, 065003 (2014).
\newblock \doi{10.1088/0954-3899/41/6/065003}

\bibitem{LlanesEstrada:2011kc}
F.J. Llanes-Estrada, O.I. Pavlova, R.~Williams, Eur. Phys. J. \textbf{C72},
  2019 (2012).
\newblock \doi{10.1140/epjc/s10052-012-2019-9}

\bibitem{SanchisAlepuz:2011aa}
H.~Sanchis-Alepuz, R.~Alkofer, G.~Eichmann, R.~Williams, PoS
  \textbf{QCD-TNT-II}, 041 (2011)

\bibitem{Qin:2018dqp}
S.X. Qin, C.D. Roberts, S.M. Schmidt, Phys. Rev. \textbf{D97}(11), 114017
  (2018).
\newblock \doi{10.1103/PhysRevD.97.114017}

\bibitem{Brown:2014ena}
Z.S. Brown, W.~Detmold, S.~Meinel, K.~Orginos, Phys. Rev. \textbf{D90}(9),
  094507 (2014).
\newblock \doi{10.1103/PhysRevD.90.094507}

\bibitem{Meinel:2012qz}
S.~Meinel, Phys. Rev. \textbf{D85}, 114510 (2012).
\newblock \doi{10.1103/PhysRevD.85.114510}

\bibitem{Briceno:2012wt}
R.A. Briceno, H.W. Lin, D.R. Bolton, Phys. Rev. \textbf{D86}, 094504 (2012).
\newblock \doi{10.1103/PhysRevD.86.094504}

\bibitem{Padmanath:2013zfa}
M.~Padmanath, R.G. Edwards, N.~Mathur, M.~Peardon, Phys. Rev. \textbf{D90}(7),
  074504 (2014).
\newblock \doi{10.1103/PhysRevD.90.074504}

\bibitem{Namekawa:2013vu}
Y.~Namekawa, et~al., Phys. Rev. \textbf{D87}(9), 094512 (2013).
\newblock \doi{10.1103/PhysRevD.87.094512}

\bibitem{Alexandrou:2014sha}
C.~Alexandrou, V.~Drach, K.~Jansen, C.~Kallidonis, G.~Koutsou, Phys. Rev.
  \textbf{D90}(7), 074501 (2014).
\newblock \doi{10.1103/PhysRevD.90.074501}

\bibitem{Can:2015exa}
K.U. Can, G.~Erkol, M.~Oka, T.T. Takahashi, Phys. Rev. \textbf{D92}(11), 114515
  (2015).
\newblock \doi{10.1103/PhysRevD.92.114515}

\bibitem{Isgur:1978wd}
N.~Isgur, G.~Karl, Phys. Rev. \textbf{D19}, 2653 (1979).
\newblock \doi{10.1103/PhysRevD.19.2653}.
\newblock [Erratum: Phys. Rev.D23,817(1981)]

\bibitem{Olsen:2017bmm}
S.L. Olsen, T.~Skwarnicki, D.~Zieminska, Rev. Mod. Phys. \textbf{90}(1), 015003
  (2018).
\newblock \doi{10.1103/RevModPhys.90.015003}

\bibitem{Brambilla:2010cs}
N.~Brambilla, et~al., Eur. Phys. J. \textbf{C71}, 1534 (2011).
\newblock \doi{10.1140/epjc/s10052-010-1534-9}

\bibitem{Brambilla:2014jmp}
N.~Brambilla, et~al., Eur. Phys. J. \textbf{C74}(10), 2981 (2014).
\newblock \doi{10.1140/epjc/s10052-014-2981-5}

\end{thebibliography}

\end{document}